\documentclass[superscriptaddress,letterpaper, amssymb,preprint, amsmath, 12pt]{revtex4-1}
 \usepackage[justification=raggedright,singlelinecheck=false,labelfont=bf,labelsep=period]{caption,subcaption}

\usepackage{braket}
\usepackage{siunitx}
\usepackage{comment}

\newcommand{\JcmK}{Jcm$^{-3}$K$^{-1}$}

\renewcommand{\thefootnote}{\fnsymbol{footnote}}

\usepackage{scalerel}
\usepackage{tikz}
\usetikzlibrary{svg.path}
\definecolor{orcidlogocol}{HTML}{A6CE39}
\tikzset{
  orcidlogo/.pic={
    \fill[orcidlogocol] svg{M256,128c0,70.7-57.3,128-128,128C57.3,256,0,198.7,0,128C0,57.3,57.3,0,128,0C198.7,0,256,57.3,256,128z};
    \fill[white] svg{M86.3,186.2H70.9V79.1h15.4v48.4V186.2z}
                 svg{M108.9,79.1h41.6c39.6,0,57,28.3,57,53.6c0,27.5-21.5,53.6-56.8,53.6h-41.8V79.1z M124.3,172.4h24.5c34.9,0,42.9-26.5,42.9-39.7c0-21.5-13.7-39.7-43.7-39.7h-23.7V172.4z}
                 svg{M88.7,56.8c0,5.5-4.5,10.1-10.1,10.1c-5.6,0-10.1-4.6-10.1-10.1c0-5.6,4.5-10.1,10.1-10.1C84.2,46.7,88.7,51.3,88.7,56.8z};}}
\newcommand\orcidicon[1]{\href{https://orcid.org/#1}{\mbox{\scalerel*{
\begin{tikzpicture}[yscale=-1,transform shape]
\pic{orcidlogo};
\end{tikzpicture}
}{|}}}}

\usepackage{etoolbox}

\makeatletter
\pretocmd\frontmatter@thefootnote{\color{blue}}{}{}

\usepackage{dcolumn} 
\usepackage{titlesec}
\titlespacing*{\section}
{0pt}{4.5ex plus 1ex minus .2ex}{1.3ex plus .2ex}
\titleformat{\section}
  {\centering\bfseries\MakeUppercase}{\thesection}{1em}{}

\usepackage{xstring}
\usepackage{xspace}
\newcommand{\SM}{Supplementary Material}
\usepackage[utf8]{inputenc}
\usepackage[english]{babel}
\usepackage{natbib}
\usepackage{gensymb}
\usepackage{mathtools}

\usepackage{float}
\usepackage{graphicx}

\usepackage{url}

\newcommand{\wmk}{Wm$^{-1}$K$^{-1}$}

\usepackage[colorlinks,citecolor=red,urlcolor=blue,bookmarks=false,hypertexnames=true]{hyperref}

\begin{document}

\title{Origin of high thermal conductivity in disentangled ultra-high molecular weight polyethylene films: ballistic phonons within enlarged crystals}

\author{Taeyong Kim~\orcidicon{0000-0003-2452-1065}\,}

\affiliation{%
Division of Engineering and Applied Science, California Institute of Technology, Pasadena, California 91125, USA
}%

\author{Stavros X. Drakopoulos~\orcidicon{0000-0002-6798-0790}\,}
\affiliation{%
Department of Materials, Loughborough University, Loughborough LE11 3TU, United Kingdom
}%

\author{Sara Ronca~\orcidicon{0000-0003-3434-6352}\,}
\affiliation{%
Department of Materials, Loughborough University, Loughborough LE11 3TU, United Kingdom
}%

\author{Austin J. Minnich~\orcidicon{0000-0002-9671-9540} }
 \email{aminnich@caltech.edu}
\affiliation{%
Division of Engineering and Applied Science, California Institute of Technology, Pasadena, California 91125, USA
}%      

\date{\today}
               
\begin{abstract}

The thermal transport properties of oriented polymers are of fundamental and practical interest. High thermal conductivities  ($\gtrsim 50$ \wmk) have recently been reported in disentangled ultra-high molecular weight  polyethylene (UHMWPE) films, considerably exceeding prior reported values for oriented films. However, conflicting explanations have been proposed for the microscopic origin of the high thermal conductivity.  Here, we report a characterization of the thermal conductivity and mean free path accumulation function of disentangled UHMWPE films (draw ratio $\sim 200$) using cryogenic steady-state thermal conductivity measurements and transient grating spectroscopy. We observe a marked dependence of the thermal conductivity on grating period over temperatures from 30 -- 300 K.  Considering this observation, cryogenic bulk thermal conductivity measurements, and analysis using an anisotropic Debye model, we conclude that longitudinal atomic vibrations with mean free paths around 400 nanometers are the primary heat carriers and that the high thermal conductivity for draw ratio  $\gtrsim 150$ arises from the enlargement of extended crystals with drawing. The  mean free paths appear to remain limited by the extended crystal dimensions, suggesting that the upper limit of thermal conductivity of disentangled UHMWPE films has not yet been realized.

\end{abstract}
\maketitle

\section{Introduction}

Thermally conductive polymers are of interest  for fundamental materials science as well as applications such as thermal management \cite{Chen_PPS_2016, Mark_Handbook_2007, Chen_RSER, Prasher_IEEEProc, TengeiZhiting_Review, Ronggui_Review}. Although the  thermal conductivity of unoriented polymers is generally less than 1 \wmk~\cite{Mark_Handbook_2007},   early works reported orders of magnitude increase in uniaxial thermal conductivity of oriented samples, including  polyethylene (PE) \cite{Hansen_WAXS_1972,BurgessGreig_1975}, polyacetylene \cite{Piraux_SSC_1989}, and polypropylene \cite{Choy_JPCSSP_1977,Choy_JPSPPE_1980}. In particular,  the reported thermal conductivity of oriented polyethylene ranged from $\sim$ 14 \wmk~\cite{Choy_Polymer_1978} for draw ratio $DR= 25$ up to $\sim$ 40 \wmk~ for solution processed PE with a DR of 350 \cite{Choy_JPS_1999}. The enhancement was attributed to various mechanisms including increased chain alignment along the drawing direction \cite{Hansen_WAXS_1972, Xinwei_2015}, phonon focusing in the elastically anisotropic crystalline phase \cite{Pietralla1989, Mergenthaler_ZPhysB}, and increased crystallinity \cite{Gibson_1977, Choy_Polymer_1978}. Recently, thermal conductivity values around $20 - 30$ \wmk~and $\gtrsim 60$~\wmk~have been reported in PE microfibers \cite{Xinwei_2015, Wang_Mamo_2013} and nanofibers  \cite{Shen_NatNano_2010, Shrestha_NatCommun_2018}, respectively. In macroscopic samples, the introduction of disentangled ultra-high molecular weight polyethylene (UHMWPE) films~\cite{Rastogi_Mamo_2011,Rastogi_nmat} with higher crystallinities and less entangled amorphous regions compared to prior samples has led to reports of high thermal conductivities exceeding 60 \wmk~\cite{Ronca_Polymer_2017, Xu_NatCommun_2019}. Several recent studies have also reported high thermal conductivity up to 20 -- 30 \wmk~ in a diverse set of polymers besides PE, including polybenzobisoxazole~\cite{Wang_Mamo_2013},  polyethylene oxide~\cite{Lu_Polymer_2017}, and amorphous polythiophene \cite{Singh_Natnano_2014}.

Knowledge of the structural changes that occur upon drawing from nascent PE aid in identifying the origin of the high thermal conductivity values, and extensive studies have characterized the atomic and nanoscale structure of PE films at different DR. The nascent structure consists of spherulites \cite{Bunn_1945} which are in turn composed of unoriented stacked lamella in which folded chains are bridged by intra- and inter-lamella tie molecules \cite{Keller_1968}. The initial crystalline fraction is on the order of $\sim 60- 70$\%, as measured using nuclear magnetic resonance (NMR) \cite{Wilson_NMR_1953} or heat capacity measurements \cite{PeterlinMeinel_Calorimetry_1965}, and from SAXS the crystalline domains have long periods  $\sim 10 - 30$ nm \cite{Fischer_1962,KOBAYASHI_SAXS_Nat1962,Yeh_1967} with the corresponding size of the crystalline domain inside the unit being around 90\% of the long period \cite{Geil_1964}. On drawing, Peterlin proposed a sequence of processes occur in which stacked lamella transition to micro-fibrils and eventually to chain-extended crystals \cite{Peterlin_theory_1971}. More precisely, on initial drawing, the lamellae begin to align and a crystalline micro-fibril structure bridged by amorphous domains or tie molecules emerges as the lamellae are fragmented. Subsequently, for $10 \lesssim DR \lesssim 50$ the micro-fibrils aggregate with concurrent tautening of the tie-molecules and marginal changes in the crystallinity. Finally, for $DR>50$, chain extension leads to an extended crystal phase formed from the aggregated micro-fibrils and tie-molecules. The density of states and dispersion of atomic vibrations in the crystalline phase have been characterized by various inelastic scattering techniques \cite{Danner_INS_1964,Safford_INS_1967, Holliday_INS_1971, Feldkamp1968,Twistleton_C11_11,Heyer_INS,Mermet_IXS_2003}.

Experimental evidence in support of the above picture has been obtained using various techniques such as transmission electron microscopy (TEM) \cite{Smith_EM_1985,BRADY_TEM_1989}, SAXS \cite{VanAerle_Xray_1988},  wide angle x-ray scattering (WAXS) \cite{VanAerle_Xray_1988}, and NMR \cite{Hu_NMR_2000,Litvinov_MaMo_2011}, among others \cite{Magonov_AFM_1993, Heyer_INS}. For instance, the formation of the micro-fibrils via lamellae fragmentation is consistent with a lack of a clear trend of crystallite size with DR for DR $\lesssim 10$ \cite{Zubov_SAXS_1992, VanAerle_Xray_1988}. The subsequent unfolding and tautening of tie-molecules along with the aggregation of microfibrils is consistent with an initial rapid increase in  crystallinity, elastic modulus and orientation factor with DR  below DR 20 \cite{Anandakumaran_Mamo_1988} followed by a marginal increase of only  a few percent up to DR as large as 200  \cite{Anandakumaran_Mamo_1988, VanAerle_Xray_1988, Stein_xray}. Evidence for the existence of extended crystal was obtained using various complementary methods such as SAXS and WAXS \cite{Litvinov_MaMo_2011, Zubov_SAXS_1992, TANG_SAXS}, TEM \cite{BRADY_TEM_1989, Smith_EM_1985}, and NMR~\cite{Hu_NMR_2000}. These various techniques indicated that the diameter of the extended crystal is $\sim 10-20$ nm and of longitudinal dimension  around 100 -- 250 nm or greater \cite{Litvinov_MaMo_2011}.

Morphological changes under drawing identified from the above structural studies have provided insight into origin of increase in thermal conductivity.  Below DR $\sim 50$, the uniaxial thermal conductivity is observed to monotonically increase with DR \cite{Choy_JPSPPE_1980, Choy_Gelspun_1993}. Various effective-medium type models have been proposed to interpret this increase in terms of changes in the crystallinity and crystalline orientation~\cite{Hennig, ChoyandYoung_1977, Takayanagi}.  Although these models are generally successful in explaining measured thermal conductivity data, the actual transport processes may differ from those assumed by  effective medium theory because of the presence of phonons that are ballistic over multiple crystallites.  Evidence of such processes has been reported even in partially oriented PE samples with low DR $\sim 7.5$ using transient grating spectroscopy (TG) \cite{ABR_PNAS_2019}.

For disentangled UHMWPE of DR $\gtrsim 150$, the increase in thermal conductivity is difficult to interpret using the above models because the thermal conductivity is observed to increase on average by factor of $\sim 20$ \% despite a lack of detectable change in crystallinity or chain orientation \cite{Choy_Polymer_1978,Choy_JPSPPE_1980, Choy_JPS_1999, Ronca_Polymer_2017, Xu_NatCommun_2019}. Conflicting explanations have been proposed to account for these observations. For instance,  Xu et al. used the isotropic helix-coil model to conclude that the thermal conductivity of the amorphous phase ($\kappa_a$) must be as high as 16 \wmk~ to explain the high thermal conductivity for samples with DR $\sim 60$ \wmk~\cite{Xu_NatCommun_2019}. On the other hand, Ronca et al. used the same model to conclude that the high thermal conductivity for $DR \gtrsim 150$ originates from the enlargement of the extended crystal dimensions~\cite{Ronca_Polymer_2017}. The discrepancy is difficult to resolve by bulk thermal conductivity measurements because the properties of the crystalline and amorphous phases cannot be independently measured. As a result, the physical origin of the high thermal conductivity of disentangled UHMWPE  remains unclear.

Here, we report measurements of the thermal conductivity and mean free path accumulation function of  disentangled UHMWPE films (DR $\sim 200$) using cryogenic thermal conductivity measurements and transient grating spectroscopy. The thermal conductivity exhibits a marked grating dependence, indicating the presence of ballistic heat-carrying atomic vibrations over the length scale of a grating period. We interpret the TG and cryogenic thermal conductivity measurements using an anisotropic Debye model that describes heat transport by longitudinal atomic vibrations. The analysis indicates that the heat is nearly entirely carried by this branch, with values of the temperature-independent mean free paths being around 400 nm up to several THz. Comparing these results to those of our prior study of disentangled UHMWPE films of lower DR, we find that the high thermal conductivity  for DR $\gtrsim 150$ can be attributed to the presence of enlarged extended crystals. As the phonon MFPs appear to be limited by the dimensions of the extended crystals, our study suggests that disentangled UHMWPE films with higher thermal conductivity may be realized in samples with larger extended crystals.

\section{Experiment}

    \begin{figure}
 {\includegraphics[width=\textwidth,height=12cm,keepaspectratio]{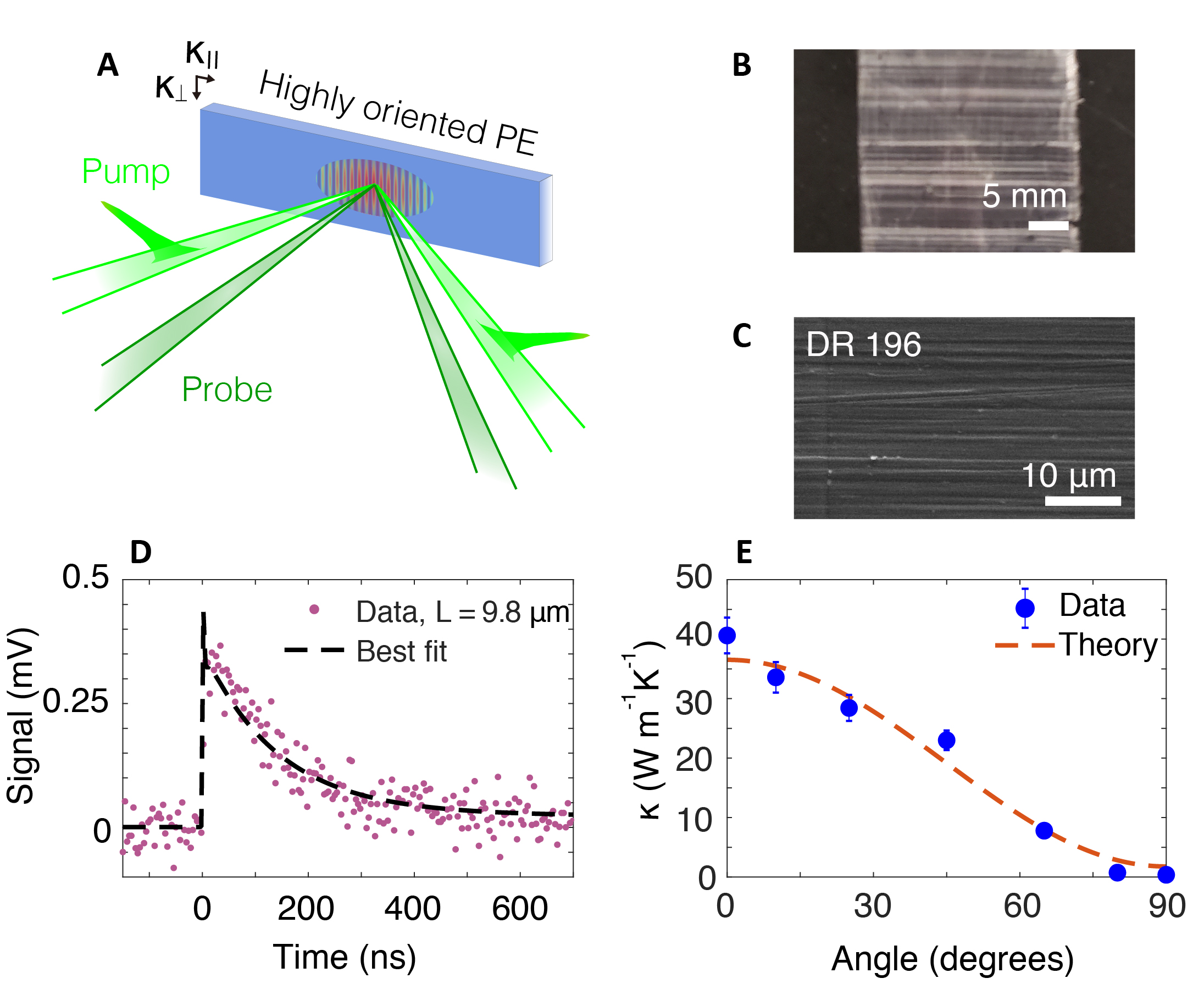}
\phantomsubcaption\label{fig:schem}
\phantomsubcaption\label{fig:optsamp}
\phantomsubcaption\label{fig:SEMsamp}
\phantomsubcaption\label{fig:repsig}
\phantomsubcaption\label{fig:angle}}
    \caption{
(A) Schematic illustration of transient grating formation and  temperature profile in disentangled UHMWPE films. Pump laser pulses impulsively generate a spatial grating on the sample from which probe beams diffract. 
(B) Optical image of disentangled UHMWPE film (DR = 196).
(C) SEM image of the disentangled UHMWPE film. Extended fibers over tens of $\mu m$ are visible.
(D) Representative TG signal versus time for  grating period $L=9.8$ $\mu m$ at 300 K. The signal is an average of 3$\times$10$^4$ repetitions at a single location on the sample. The measurement was conducted at multiple locations (see \SM~Sec.~1 for additional data). The thermal diffusivity is obtained as the time constant of the exponential decay.
(E) Thermal conductivity versus angle between draw direction and thermal gradient defined by the grating  $L = 9.8$ $\mu$m. The 0\degree~ (90\degree) indicates heat flow direction parallel  (perpendicular) to the draw direction. The maximum thermal conductivity is around 40 \wmk~ along the chain, while the perpendicular value is comparable to that of unoriented PE.}
\end{figure}

We measured the in-plane thermal conductivity of disentangled UHMWPE films using TG, as schematically illustrated in Fig.~\ref{fig:schem}. The TG setup is identical to that described in Ref.~\cite{ABR_PNAS_2019}. Briefly, a pair of pump pulses (wavelength 515 nm, beam diameter 530 $\mu$m, pulse duration $\sim$ 1 ns, pulse energy 13 $\mu$J, repetition rate 200 Hz) is focused onto the sample to impulsively create a spatially sinusoidal temperature rise of period $L$ and wave vector $q=2 \pi/L$. The grating relaxes by heat conduction, and its decay is monitored by the heterodyne measurement of a diffracted CW signal beam and reference probe beam  (wavelength 532 nm, beam diameter 470 $\mu$m, CW power 900 $\mu$W, chopped at 3.2 \% duty cycle to reduce steady heating on the sample). The samples are disentangled UHMWPE films synthesized using the same procedure given in Ref.~\cite{Rastogi_Mamo_2011}, but with higher draw ratio ($DR = 196$) achieved by rolling  ($\times 7$) and stretching ($\times 28$). Figure~\ref{fig:optsamp} shows an optical image of the film that is of centimeter scale dimension laterally and of thickness around 30 micrometers, as measured using calipers. A scanning electron microscope (SEM) image is given in Fig.~\ref{fig:SEMsamp}. In both images, highly oriented fibers extending over tens of microns are visible. Since PE is transparent to visible light, Au nanoparticles (diameter:  $\sim 2 - 12$ nm \cite{Stavros_AuSize}; concentration: 1 wt\%)  were added as an optical absorber. The concentration of the Au was selected to minimize the effect of the filler on the thermal conductivity while enabling the formation of a thermal grating on the sample \cite{ABR_PNAS_2019}. Experimental characterization of similar samples using polarized light microscopy, among other methods, indicates that the nanoparticles are oriented in linear chains in the amorphous regions. \cite{Stavros_AuSize}

 A representative TG signal measured at grating period $L = 9.8$ $\mu m$ is shown in Fig.~\ref{fig:repsig}. As described in the Supporting Information of Ref.~\cite{ABR_PNAS_2019}, the signal consists of an initially fast decay (time constant $\lesssim 1$ ns) followed by a slower decay (time constant $\gtrsim 10$ ns).  The initial fast decay is  attributed to the thermal relaxation of the Au nanoparticles, while the subsequent slower decay corresponds to thermal conduction in the film. Following the procedure in Ref.~\cite{ABR_PNAS_2019}, we fit the signal with a multi-exponential function; the time constant of the slower decay yields the thermal diffusivity of the sample.  Because the initial signal from the nanoparticles exhibits a short time constant compared to their thermal signal, the influence of the nanoparticle signal on the fitted thermal diffusivity is negligible. The signal-to-noise ratio (SNR) of the present measurements is generally less than 20,  which is about 30\% of that reported in Ref.~\cite{ABR_PNAS_2019} for $DR =7.5$ samples . This decrease is because the highly oriented samples scatter visible light intensely owing to the increased inhomogeneity over length scales comparable to the optical wavelength, as evidenced by  AFM images perpendicular to the fiber alignment direction (see \SM~Sec.~4). Nevertheless, TG is able to measure the thermal signal with adequate SNR because only the diffracted signal due to the spatial refractive index profile at the grating wave vector is measured, and the scattered light intercepted by the detector that does not arise from diffraction can be subtracted from the final signal using a heterodyning procedure \cite{Jeremy_PRL_2013}.  The thermal conductivity was calculated from the measured thermal diffusivity using the heat capacities of linear PE in Ref.~\cite{Chang_heatcap}. 
 
 The in-plane  thermal conductivity versus angle between the fiber alignment direction and the thermal gradient is shown in Fig.~\ref{fig:angle}. The thermal conductivity is $\sim$ 40 \wmk~at 0\degree ~($\kappa_{\parallel}$, parallel to grating), and decreases to 0.4 \wmk~at 90\degree~($\kappa_{\bot}$, perpendicular to grating). The value along the draw direction is in reasonable agreement with that obtained on a sample without Au nanoparticles using the laser flash method \cite{Ronca_Polymer_2017}.  The value of $\kappa_{\bot} \sim 0.4$ \wmk~is close to that of unoriented PE, which is attributed to the heat conduction by interchain van der Waals interactions  \cite{Choy_polymer_1977}. The angle-dependent thermal conductivity was fitted by a geometric model \cite{Zolotoyabko_Book} with the thermal conductivity along the two principal directions as input. Good agreement between the model and the data is observed.

   \begin{figure}
    \centering
{\includegraphics[width=\textwidth,height=\textheight/2,keepaspectratio]{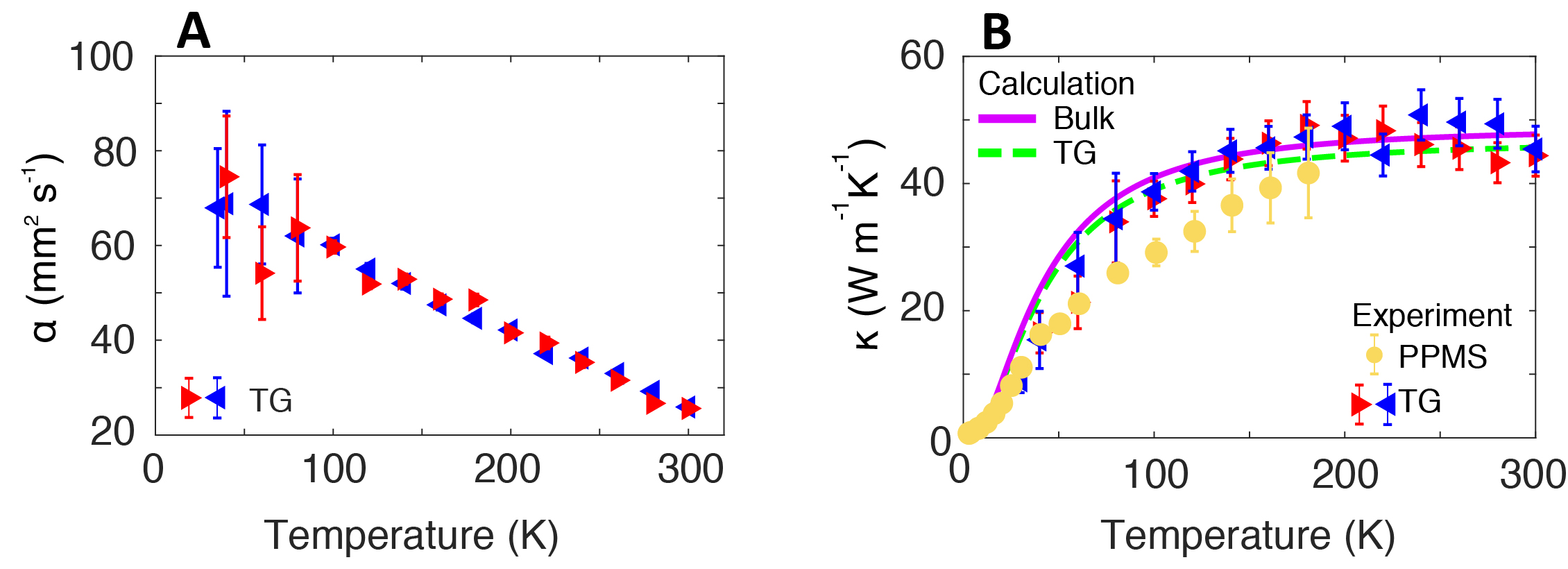}
\phantomsubcaption\label{fig:diffusivity}
\phantomsubcaption\label{fig:kappavsT}}
\caption{
(A) Thermal diffusivity along the chain axis versus  temperature for grating
period $L = 9.8$ $\mu m$. An increase in the thermal
diffusivity is observed as temperature decreases.
(B) Thermal conductivity along the chain axis versus  temperature measured by TG for $L = 9.8$ $\mu m$ (blue triangles: cooling, red triangles: heating)
) and PPMS (yellow circles).  The thermal conductivity is approximately constant from  $\sim 300-220$ K, below which the thermal conductivity decreases. The trend and corresponding values from PPMS and TG with $L=9.8$ $\mu m$ are in reasonable agreement, suggesting that phonon mean free paths are less than $\sim L / 2\pi \sim 1.5$ micrometers. Calculated thermal conductivity versus temperature obtained using  Eq.~\ref{eq:kappaiani} (solid purple line: bulk; dashed green line: $L=9.8$ $\mu m$). 
}
\end{figure}

The thermal transport properties can be further examined by measuring the temperature dependence of the thermal diffusivity and conductivity. The bulk thermal diffusivity along the chain direction versus temperature obtained from TG with $L=9.8$ micrometers between 30 -- 300 K is shown in Fig.~\ref{fig:diffusivity}. The diffusivity exhibits a linearly increasing trend with decreasing temperature, a qualitatively similar trend as that reported in microscale crystalline fibers \cite{Xinwei_2015}. The corresponding $\kappa_{\parallel}$  versus temperature  is shown in Fig.~\ref{fig:kappavsT}. Within the uncertainty of the measurement, the thermal conductivity remains constant from room temperature to $\sim 220$ K, below which the thermal conductivity decreases. 

\begin{figure}
{
\includegraphics[width=\textwidth,height=\textheight/2,keepaspectratio]{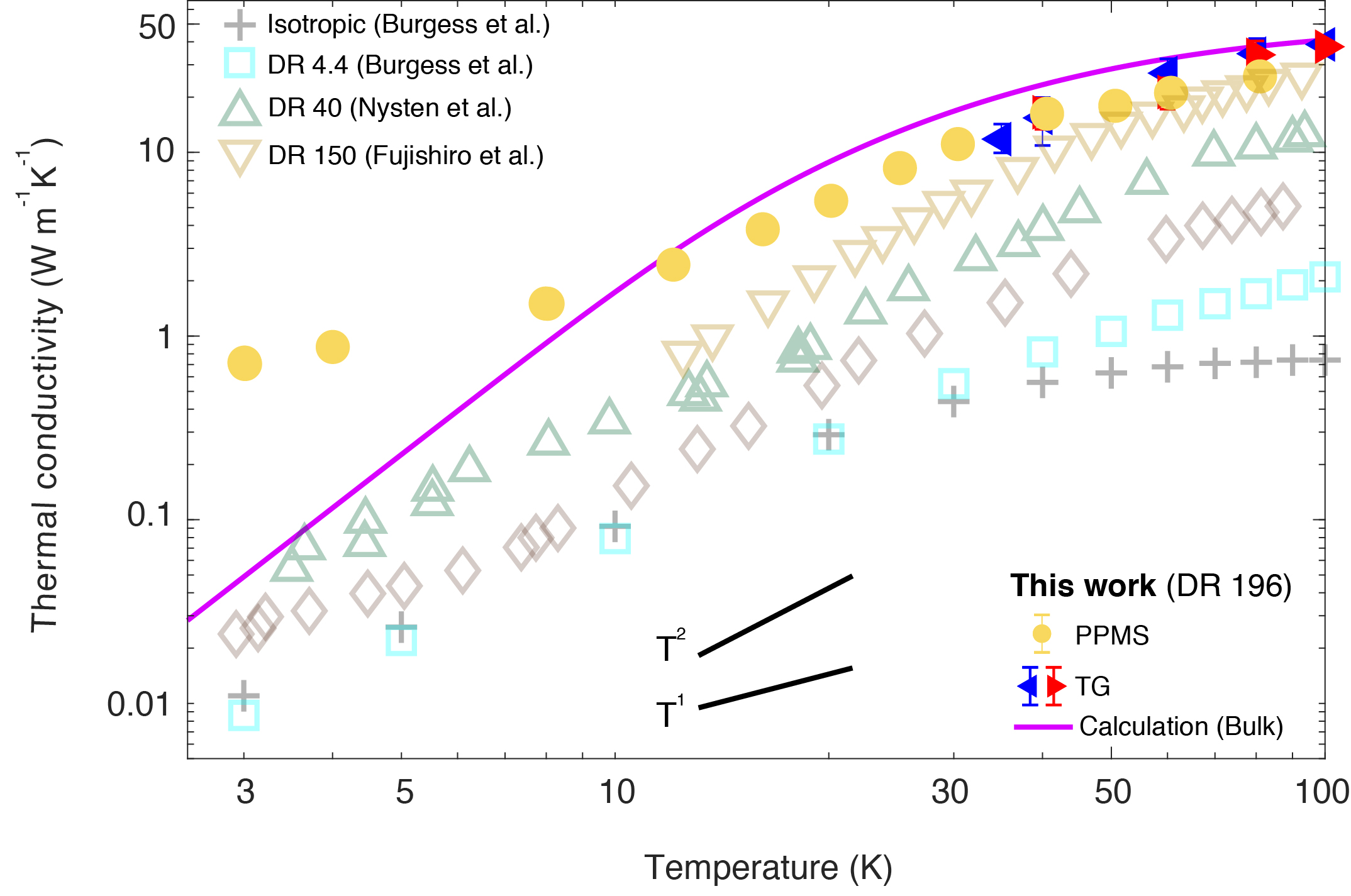}
\caption{\label{fig:LowTkappa}
Cryogenic bulk thermal conductivity versus temperatures below 100 K from TG (filled colored triangles), PPMS (yellow circles) and calculated (purple solid line). Representative  literature data of thermal conductivity versus  temperature for semi-crystalline PE with various DRs are also plotted as open symbols (extruded thin film, DR 4.4,  Ref.~\cite{BurgessGreig_1975}; solution-cast thin film, DR 40, Ref.~\cite{NYSTEN_1995}; solution-cast macroscopic fiber, DR 150, in Ref.~\cite{Fujishiro_1998}. As  temperature decreases, the trend of   measured thermal conductivity exhibits a transition from $\sim T^2$ to $\sim T$ near 10 K. }%This trend is evident in the other data except that of the isotropic film.**}
}
\end{figure}

The macroscopic dimensions of the present samples permits additional characterization of the thermal conductivity down to $\sim 3$ K using a  Physical Property Measurement System (PPMS) (See \SM~Sec.~3 for further details). The results are shown in Fig.~\ref{fig:kappavsT}. The measured values and the trend of the bulk thermal conductivity are consistent with that obtained from TG. The cryogenic thermal conductivity values on a logarithmic scale are given in Fig.~\ref{fig:LowTkappa}. The values exhibit two distinct temperature dependencies with a transition at around 10 K. 

The temperature dependence of the  thermal diffusivity and conductivity provides insight into the origin of the high thermal conductivity in the present samples. First, the thermal diffusivity is observed to depend on temperature, ruling out a constant relaxation time for all phonon polarizations as suggested in Ref.~\cite{Shrestha_NatCommun_2018}.  Second,  the measured trend of  thermal conductivity versus  temperature is consistent with structural scattering being the dominant scattering mechanism. Above 10 K, the trend is generally consistent with  those of previously reported data on PE films of various DR as shown in Fig.~\ref{fig:LowTkappa}, although the thermal conductivity of the present sample is consistently higher. Below 10 K, a weaker trend with temperature is observed compared those exhibited by other samples in the same temperature range. This difference may be due to the disentangled nature of the present sample compared to prior solution-cast films and will be the topic of future study. Third, within the uncertainty of the measurements, the measured bulk thermal conductivity is in reasonable agreement with the TG data for $L=9.8$ $\mu m$, indicating phonon mean free paths are shorter than $\sim 9.8/2\pi \sim 1.5$ micrometers ~\cite{Austin_Determining}.

\begin{figure}
{\includegraphics[width=.7\textwidth,height=\textheight,keepaspectratio]{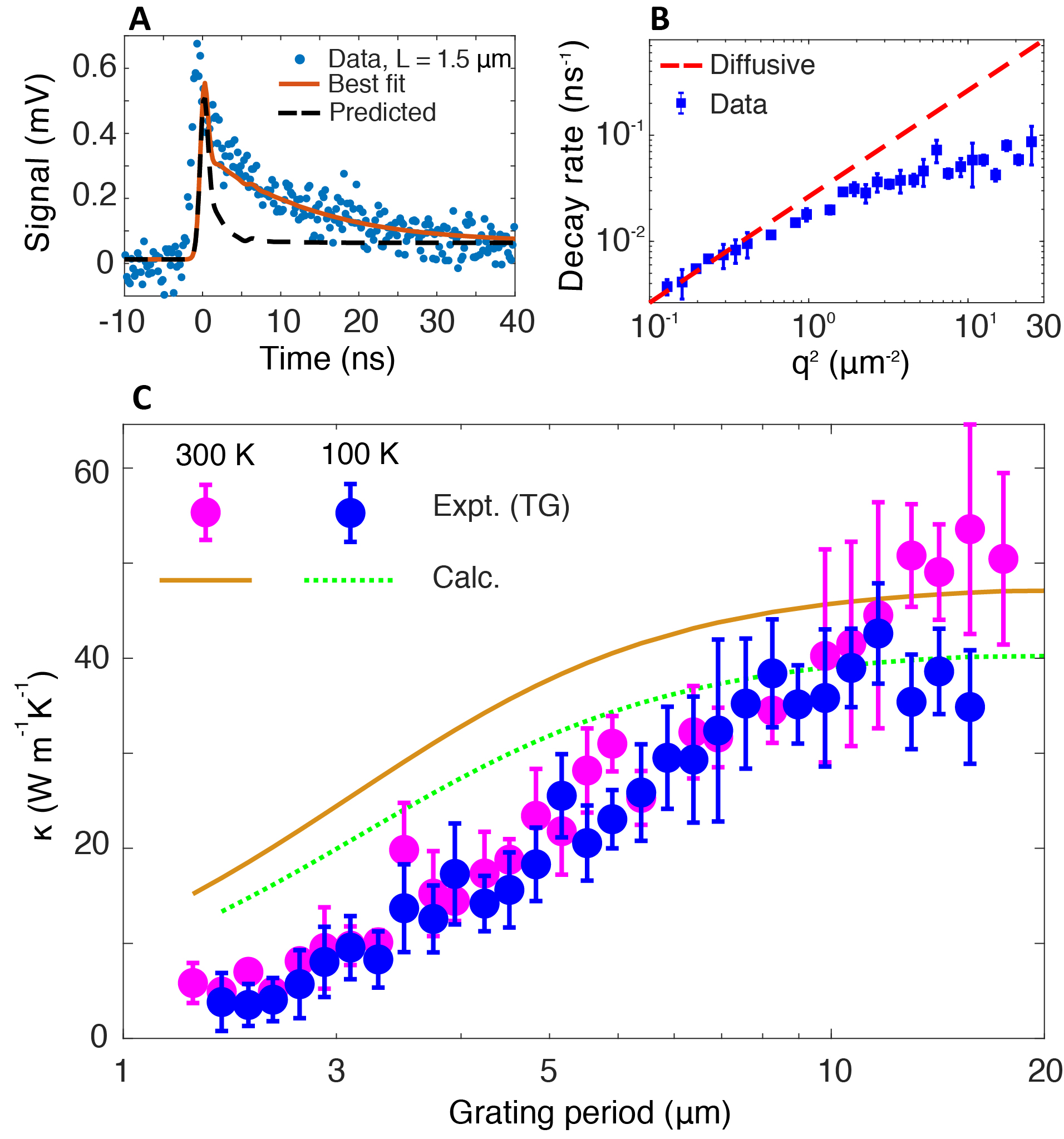}
\phantomsubcaption\label{fig:TGsigL1.5}
\phantomsubcaption\label{fig:gammavsqsq}
\phantomsubcaption\label{fig:grating}}
\caption{\label{fig:Ballistic}
(A) Measured TG signal versus time (symbols) for grating period $L = 1.5~\mu m$, corresponding to $q^2 \sim 17.5$ $\mu m^{-2}$, along with the best fit (solid red line) and predicted decay estimated using the thermal conductivity obtained for $L=9.8$ $\mu$m (dashed black line). The signal is an average of $5 \times 10^4$ shots measured at a single location.  The signal clearly decays slower than predicted based on the thermal conductivity measured at larger grating period, indicating a departure from  diffusive thermal transport.
(B) TG signal decay rate versus $q^2$. The measured decay rates for $q^2 \gtrsim 0.3$ $\mu m^{-2}$ deviate from that predicted from the thermal conductivity measured at $L=9.8$ micrometers.
(C) Experimental thermal conductivity versus grating period at selected temperatures from experiments (magenta symbols: 300 K; blue symbols: 100 K) along with the calculation (orange solid line: 300 K; green dotted line: 100 K).  The thermal conductivity exhibits a marked dependence on grating period and is nearly independent of temperature.  Error bars indicate 95\% confidence intervals  obtained using the procedure given in Ref.~\cite{ABR_PNAS_2019}. }
\end{figure}

Despite these constraints from the temperature-dependent bulk thermal transport properties, the microscopic properties of the heat-carrying atomic vibrations remain underdetermined. To gain further insight, we exploited the ability of TG to systematically vary the induced thermal gradient over micrometer length scales by tuning the grating period. If heat-carrying phonons propagate ballistically over the grating period, the thermal decay is slower than that predicted from the bulk thermal conductivity \cite{Jeremy_PRL_2013, Austin_Determining}. We have previously used this approach to identify ballistic phonons over nanocrystalline domains in disentangled UHMWPE samples of lower DR \cite{ABR_PNAS_2019}.

We apply this approach to the present samples, measuring the thermal conductivity along the chain axis versus  grating period at temperatures of 300 K, 220 K, 100 K, and 35 K. The measured TG signal for a grating period of $L = 1.5$ $\mu m$ is shown in the inset of Fig~\ref{fig:TGsigL1.5}. The decay is clearly slower than  expected based on the bulk thermal conductivity value, indicating the presence of ballistic phonons on the length scale of the grating period. Measurements of the decay rate versus $q^2$ for all the grating periods at 300 K are given in  Fig. \ref{fig:gammavsqsq}. The measured decay rate is close to that  predicted by the bulk thermal conductivity at $q^2 \leq 0.3$ $\mu m^{-2}$, above which the decay rate is slower by up to a factor of 5 at the smallest grating period.

The corresponding thermal conductivity versus grating period at 300 K and 100 K is shown in Fig.~\ref{fig:grating}. The thermal conductivity exhibits a marked dependence on grating period up to $\sim 10$  $\mu$m, a value comparable to that observed in other covalent crystals  with higher thermal conductivity such as silicon \cite{Navaneeth_PRX_2018, Jeremy_PRL_2013}. Compared to PE samples of lower DR $\sim 7.5$~\cite{ABR_PNAS_2019}, the observed trend is significantly more pronounced, indicating the presence of heat-carrying phonons with longer MFPs in the present  samples. The observed grating dependence lacks a clear temperature dependence for the temperatures considered here (see~\SM~Sec.~3 for additional data). This finding indicates the dominance of the structural scattering, consistent with the trend of the bulk thermal conductivity versus temperature in Fig~\ref{fig:kappavsT}.

\section{Low energy Debye model}
We now construct a model to interpret the measurements in Figs~\ref{fig:kappavsT},~\ref{fig:LowTkappa}, and \ref{fig:Ballistic}. We begin by noting that Fig.~\ref{fig:Ballistic} indicates that phonons with MFPs on the order of hundreds of nanometers carry the majority of the heat. Phonons with such long MFPs are likely from the LA branch owing to its high group velocity $v_{c} \sim 16 - 17$ km s$^{-1}$ \cite{Mermet_IXS_2003, Holliday_INS_1971, Twistleton_C11_11,Feldkamp1968, Pietralla1996}. Therefore, the marked dependence of grating period in Fig.~\ref{fig:Ballistic} implies that nearly all heat is carried by the LA branch. Further, the grating-period dependent thermal conductivity in Fig.~\ref{fig:grating} does not exhibit a temperature dependence, indicating that the MFPs are independent of temperature. We  therefore construct an anisotropic Debye model ~\cite{Bowman_elasticcont1958,Dames_PRB} for the heat conducted by this branch and use the data to constrain the frequency-dependence of the LA branch relaxation time. The marked elastic anisotropy of PE can be accounted for to good approximation by assuming the group velocities all point along the chain axis \cite{Pietralla1989}. Therefore, the  thermal conductivity measured in TG ($\kappa_i$) can be expressed as

\begin{equation}
\kappa_{i}=  \int_{0}^{\omega_{ab}} S(x_{i}) \left[ C_1(\omega) v_c \Lambda(\omega) \right] d\omega+\int_{\omega_{ab}}^{\omega_{c}} S(x_{i}) \left[ C_2(\omega) v_c \Lambda(\omega) \right] d\omega 
\label{eq:kappaiani}
\end{equation}

where $q_i = 2 \pi L_i^{-1}$, $x_{i,s} = q_i \Lambda_s (\omega) $, $S(x_{i,s})$ is the anisotropic phonon suppression function for an arbitrary phonon dispersion  ~\cite{Austin_layaniso},  $C_1(\omega)$ and $C_2(\omega)$ refer to the heat capacity terms in Eq.~11b of  Ref.~\cite{Dames_PRB}.

We now specify the numerical values for this model. The c-axis velocity of the longitudinal polarization is $v_c \sim 17$ km s$^{-1}$~\cite{Holliday_INS_1971, Feldkamp1968,Twistleton_C11_11}. The velocity along a perpendicular crystal axis was reported as $v_{ab} \sim 1.35$ km s$^{-1}$ \cite{Holliday_INS_1971} from inelastic neutron scattering and $\sim 4.5$ km s$^{-1}$ from an ab-initio calculation  \cite{Nina_PRL_2017} which may be reflective of the monocrystals present in disentangled UHMWPE \cite{Rastogi_Mamo_2011}. We roughly estimate $v_{ab} \sim 3$ km s$^{-1}$. We consider $\omega_c$ to be a characteristic frequency at which the $c$-axis longitudinal velocity decreases below the Debye velocity and ultimately tends towards zero. Roughly, we estimate $\omega_c \sim 10$ THz; the analysis below is not sensitive to this choice. This choice then determines $\omega_{ab} = 1.8$ THz. The corresponding maximum wave vector magnitude is $\sim 6$ \AA$^{-1}$.

We now seek to identify the function $\Lambda (\omega)$ that best explains the temperature  and grating dependence of the thermal conductivity. To constrain the MFP function, we note that at $\sim 1$ THz,  the magnitude of the MFP can be  estimated using the dominant phonon approximation \cite{Zeller_DPA} and the cryogenic thermal conductivity measurements in Fig~\ref{fig:LowTkappa}. We use the cryogenic thermal conductivity at around 12 K, the minimum temperature that is still on the $T^2$ trend of the thermal conductivity. We find $\Lambda  \sim \kappa/C_l v_l \sim 340$ nm at $\sim 1$ THz, where $\kappa\sim 2.45$~\wmk~and $C_l\sim 4.2\times10^{-4}$~\JcmK~is the computed heat capacity of the longitudinal branch at 12 K.

We now consider various MFP profiles versus frequency that yield the best agreement with the experiments in Figs.~\ref{fig:kappavsT},~\ref{fig:LowTkappa}, and \ref{fig:Ballistic} by adjusting the MFP profile subject to the above constraint.  After extensive comparison, we found that the best fit is obtained using a constant $\Lambda \sim 350 - 400$ nm up to $\sim 4.5 - 6$ THz, beyond which the MFP decreases as $\sim \omega^{-4}$, although similar power laws also yield reasonable agreement.  (See \SM~Sec.~3 for further details on results using other candidate profiles).

The resulting computed bulk thermal conductivity using this profile is presented in Figs.~\ref{fig:kappavsT} and \ref{fig:LowTkappa}. In Fig.~\ref{fig:kappavsT}, the bulk thermal conductivity exhibits qualitative agreement with the measurements, producing the observed magnitude and trend. The calculated thermal conductivity in TG with $L=9.8$ $\mu$m is close to bulk value,  consistent with the good agreement between the TG and PPMS measurements. The grating period dependence of the thermal conductivity is also qualitatively reproduced by the calculation in Fig.~\ref{fig:grating}, and Fig.~S2 in \SM~Sec.~2, although quantitative disrepancies exist.

The calculated cryogenic thermal conductivity is shown in Fig.~\ref{fig:LowTkappa}. Although a similar qualitative trend is observed above $\sim 10$ K, the agreement is worse below this temperature. This discrepancy could be attributed to heat conduction by other types of atomic vibrations which may make the primary contribution to thermal conductivity below $\sim 10$ K.

\section{Discussion}

We now discuss our findings in context with prior studies of thermal conduction in oriented PE films. We first consider the MFP value of $\sim 400$ nm for frequencies in the THz range. For frequencies around 1 THz, prior values inferred from literature data using the dominant phonon approximation at cryogenic temperatures are $\sim 80$ nm using $\kappa\sim 0.34$ \wmk~for DR 40 \cite{NYSTEN_1995}, as estimated using LA specific heat of $\sim 2.8 \times 10^{-4}$~\JcmK~at 10 K using the model described above. This value is comparable to an estimated value of $\sim 60$ nm for DR 6 at 10 K \cite{Mergenthaler_ZPhysB}.  At higher frequencies $\sim 6$ THz, IXS has been used to obtain a MFP of $\sim 50$ nm for DR 5.5 \cite{Mermet_IXS_2003}, which is in reasonable agreement with the values inferred from transport studies. The MFPs obtained from TG measurements on a DR 7.5 sample in our previous study ranged from  10 - 200 nm \cite{ABR_PNAS_2019}. All of these values are comparable to but distinctly smaller than the present value,  expected as the present samples exhibit higher thermal conductivity. The decrease in MFP at frequencies exceeding $\sim 7$ THz is also consistent with the increase in broadening reported in IXS \cite{Mermet_IXS_2003}, although this increase occurred at higher frequencies in their data.

An average MFP for all LA phonons of $\sim 7$ nm was estimated from a study of PE microfibers \cite{Wang_Mamo_2013}. This value is considerably smaller than those above; as noted in that work, this value is an underestimate owing to the use of the total heat capacity which is greater than that of the longitudinal polarization only. Finally, an ab-initio study of PE predicted MFPs of up to 200 nm \cite{Nina_PRL_2017}. This prediction thus appears to be an underestimate and is likely due to insufficient grid density to sample the stiff LA branch.  This observation implies that the predicted thermal conductivity in that work is also an underestimate of the upper limit of thermal conductivity in crystalline PE.

Our findings  help to explain the origin of high thermal conductivity in disentangled UHMWPE films. Recent studies have provided conflicting explanations for the high thermal conductivity, with Xu et al. \cite{Xu_NatCommun_2019} attributing it to the high thermal conductivity of the amorphous phase \cite{ABR_PNAS_2019} but Ronca et al. \cite{Ronca_Polymer_2017} attributing it to increased extended crystal dimensions. Our data and analysis are consistent with the latter explanation. Compared to the MFPs of a DR 7.5 sample in our prior study,\cite{ABR_PNAS_2019} the present MFPs are clearly larger, as would be expected if the extended crystal dimensions have increased. The MFPs in both samples exhibit a clear maximum in the low THz frequencies and are independent of temperature, suggesting  phonon scattering is predominantly due to reflections at crystalline domain boundaries. Further, evidence exists  for the presence of  extended crystals with length on the order of hundreds of nanometers from NMR and TEM~\cite{Hu_NMR_2000,Litvinov_MaMo_2011,BRADY_TEM_1989}, values that are compatible with the MFPs obtained here. We infer that heat conduction in disentangled UHMWPE is primarily due to longitudinal atomic vibrations that are ballistic within the extended crystal phase, scattered primarily by reflections at the boundaries between the crystals. Our results therefore support the hypothesis of Ref.~\cite{Ronca_Polymer_2017} in which the  high thermal conductivity compared to the prediction including only crystallinity and anisotropy factors for DR $\gtrsim 180$~(Fig.~7 in Ref.~\cite{Ronca_Polymer_2017}) was attributed to the enlargement of extended crystal dimensions.

 Finally, we discuss the implications of our findings for realizing PE films of higher thermal conductivity. Because the MFPs appear to be limited by the size of the extended crystals, our study indicates that the thermal conductivity of PE films has not yet reached its upper limit. The practical challenge is synthesizing disentangled UHMWPE films with larger extended crystal dimensions. Such films would be expected to have higher thermal conductivity of an amount proportional to the increase in crystalline dimension.

\section{Summary}
In summary, we have characterized the thermal conductivity and mean free path accumulation function of disentangled UHMWPE films with draw ratio of nearly 200 at various temperatures. We find that heat in PE films is nearly entirely carried by longitudinal atomic vibrations with MFPs in the sub-micron range and limited by the dimensions of the extended crystal phase. The high thermal conductivity at $DR \gtrsim 150$ exceeding the expected value considering orientation and crystallinity can be attributed to increased dimensions of the extended crystals. Considering the MFPs appear to remain limited by extended crystal dimensions, our work indicates that the thermal conductivity of disentangled UHMWPE films has not yet reached its upper limit.

\begin{acknowledgments}
The authors thank Bolin Liao and Wenkai Ouyang for  assistance with PPMS measurements.
This work was supported by the Office of Naval Research under Grant Number N00014-18-1-2101.
\end{acknowledgments}

\bibliographystyle{unsrtnat}

\bibliography{PolymerRef}

\begin{thebibliography}{72}
\providecommand{\natexlab}[1]{#1}
\providecommand{\url}[1]{\texttt{#1}}
\expandafter\ifx\csname urlstyle\endcsname\relax
  \providecommand{\doi}[1]{doi: #1}\else
  \providecommand{\doi}{doi: \begingroup \urlstyle{rm}\Url}\fi

\bibitem[Chen et~al.(2016{\natexlab{a}})Chen, Ginzburg, Yang, Yang, Liu, Huang,
  Du, and Chen]{Chen_PPS_2016}
Hongyu Chen, Valeriy~V. Ginzburg, Jian Yang, Yunfeng Yang, Wei Liu, Yan Huang,
  Libo Du, and Bin Chen.
\newblock Thermal conductivity of polymer-based composites: {Fundamentals} and
  applications.
\newblock \emph{Progress in Polymer Science}, 59:\penalty0 41--85, August
  2016{\natexlab{a}}.
\newblock ISSN 00796700.
\newblock \doi{10.1016/j.progpolymsci.2016.03.001}.
\newblock URL
  \url{https://linkinghub.elsevier.com/retrieve/pii/S0079670016000216}.

\bibitem[Mark(2007)]{Mark_Handbook_2007}
James~E. Mark, editor.
\newblock \emph{Physical {Properties} of {Polymers} {Handbook}}.
\newblock Springer-Verlag, New York, 2 edition, 2007.
\newblock ISBN 978-0-387-31235-4.
\newblock \doi{10.1007/978-0-387-69002-5}.
\newblock URL \url{https://www.springer.com/gp/book/9780387312354}.

\bibitem[Chen et~al.(2016{\natexlab{b}})Chen, Su, Reay, and Riffat]{Chen_RSER}
Xiangjie Chen, Yuehong Su, David Reay, and Saffa Riffat.
\newblock Recent research developments in polymer heat exchangers – a review.
\newblock \emph{Renewable and Sustainable Energy Reviews}, 60:\penalty0
  1367--1386, 2016{\natexlab{b}}.
\newblock ISSN 1364-0321.
\newblock \doi{https://doi.org/10.1016/j.rser.2016.03.024}.
\newblock URL
  \url{https://www.sciencedirect.com/science/article/pii/S1364032116002598}.

\bibitem[Prasher(2006)]{Prasher_IEEEProc}
R.~Prasher.
\newblock Thermal interface materials: Historical perspective, status, and
  future directions.
\newblock \emph{Proceedings of the IEEE}, 94\penalty0 (8):\penalty0 1571--1586,
  2006.
\newblock \doi{10.1109/JPROC.2006.879796}.

\bibitem[Wei et~al.(2021)Wei, Wang, Tian, and Luo]{TengeiZhiting_Review}
Xingfei Wei, Zhi Wang, Zhiting Tian, and Tengfei Luo.
\newblock {Thermal Transport in Polymers: A Review}.
\newblock \emph{Journal of Heat Transfer}, 143\penalty0 (7), 04 2021.
\newblock ISSN 0022-1481.
\newblock \doi{10.1115/1.4050557}.
\newblock URL \url{https://doi.org/10.1115/1.4050557}.
\newblock 072101.

\bibitem[Huang et~al.(2018)Huang, Qian, and Yang]{Ronggui_Review}
Congliang Huang, Xin Qian, and Ronggui Yang.
\newblock Thermal conductivity of polymers and polymer nanocomposites.
\newblock \emph{Materials Science and Engineering: R: Reports}, 132:\penalty0
  1--22, 2018.
\newblock ISSN 0927-796X.
\newblock \doi{https://doi.org/10.1016/j.mser.2018.06.002}.
\newblock URL
  \url{https://www.sciencedirect.com/science/article/pii/S0927796X1830113X}.

\bibitem[Hansen and Bernier(1972)]{Hansen_WAXS_1972}
D.~Hansen and G.~A. Bernier.
\newblock Thermal conductivity of polyethylene: The effects of crystal size,
  density and orientation on the thermal conductivity.
\newblock \emph{Polymer Engineering \& Science}, 12\penalty0 (3):\penalty0
  204--208, 1972.
\newblock \doi{10.1002/pen.760120308}.
\newblock URL
  \url{https://onlinelibrary.wiley.com/doi/abs/10.1002/pen.760120308}.

\bibitem[Burgess and Greig(1975)]{BurgessGreig_1975}
S~Burgess and D~Greig.
\newblock The low-temperature thermal conductivity of polyethylene.
\newblock \emph{Journal of Physics C: Solid State Physics}, 8\penalty0
  (11):\penalty0 1637--1648, jun 1975.
\newblock \doi{10.1088/0022-3719/8/11/015}.
\newblock URL \url{https://doi.org/10.1088/0022-3719/8/11/015}.

\bibitem[Piraux et~al.(1989)Piraux, Kinany-Alaoui, Issi, Begin, and
  Billaud]{Piraux_SSC_1989}
L~Piraux, M~Kinany-Alaoui, J.~P Issi, D~Begin, and D~Billaud.
\newblock Thermal conductivity of an oriented polyacetylene film.
\newblock \emph{Solid State Communications}, 70\penalty0 (4):\penalty0
  427--429, March 1989.
\newblock ISSN 0038-1098.
\newblock \doi{10.1016/0038-1098(89)91073-9}.
\newblock URL
  \url{http://www.sciencedirect.com/science/article/pii/0038109889910739}.

\bibitem[Choy and Greig(1977)]{Choy_JPCSSP_1977}
C~L Choy and D~Greig.
\newblock The low temperature thermal conductivity of isotropic and oriented
  polymers.
\newblock \emph{Journal of Physics C: Solid State Physics}, 10\penalty0
  (2):\penalty0 169--179, jan 1977.
\newblock \doi{10.1088/0022-3719/10/2/005}.
\newblock URL \url{https://doi.org/10.1088/0022-3719/10/2/005}.

\bibitem[Choy et~al.(1980)Choy, Chen, and Luk]{Choy_JPSPPE_1980}
C.~L. Choy, F.~C. Chen, and W.~H. Luk.
\newblock Thermal conductivity of oriented crystalline polymers.
\newblock \emph{Journal of Polymer Science: Polymer Physics Edition},
  18\penalty0 (6):\penalty0 1187--1207, 1980.
\newblock \doi{10.1002/pol.1980.180180603}.
\newblock URL
  \url{https://onlinelibrary.wiley.com/doi/abs/10.1002/pol.1980.180180603}.

\bibitem[Choy et~al.(1978)Choy, Luk, and Chen]{Choy_Polymer_1978}
C.~L Choy, W.~H Luk, and F.~C Chen.
\newblock Thermal conductivity of highly oriented polyethylene.
\newblock \emph{Polymer}, 19\penalty0 (2):\penalty0 155--162, February 1978.
\newblock ISSN 0032-3861.
\newblock \doi{10.1016/0032-3861(78)90032-0}.
\newblock URL
  \url{http://www.sciencedirect.com/science/article/pii/0032386178900320}.

\bibitem[Choy et~al.(1999)Choy, Wong, Yang, and Kanamoto]{Choy_JPS_1999}
C.~L. Choy, Y.~W. Wong, G.~W. Yang, and Tetsuo Kanamoto.
\newblock Elastic modulus and thermal conductivity of ultradrawn polyethylene.
\newblock \emph{Journal of Polymer Science Part B: Polymer Physics},
  37\penalty0 (23):\penalty0 3359--3367, 1999.
\newblock ISSN 1099-0488.
\newblock
  \doi{10.1002/(SICI)1099-0488(19991201)37:23<3359::AID-POLB11>3.0.CO;2-S}.
\newblock URL
  \url{https://onlinelibrary.wiley.com/doi/abs/10.1002/%28SICI%291099-0488%2819991201%2937%3A23%3C3359%3A%3AAID-POLB11%3E3.0.CO%3B2-S}.

\bibitem[Liu et~al.(2015)Liu, Xu, Cheng, Xu, and Wang]{Xinwei_2015}
Jing Liu, Zaoli Xu, Zhe Cheng, Shen Xu, and Xinwei Wang.
\newblock Thermal conductivity of ultrahigh molecular weight polyethylene
  crystal: Defect effect uncovered by 0 k limit phonon diffusion.
\newblock \emph{ACS Applied Materials \& Interfaces}, 7\penalty0 (49):\penalty0
  27279--27288, 2015.
\newblock \doi{10.1021/acsami.5b08578}.
\newblock URL \url{https://doi.org/10.1021/acsami.5b08578}.
\newblock PMID: 26593380.

\bibitem[Pietralla et~al.(1989)Pietralla, Weeger, and
  Mergenthaler]{Pietralla1989}
M.~Pietralla, R.~M. Weeger, and D.~B. Mergenthaler.
\newblock The role of phonon focussing and structure scattering in oriented
  semicrystalline polymers.
\newblock \emph{Zeitschrift f{\"u}r Physik B Condensed Matter}, 77\penalty0
  (2):\penalty0 219--228, Jun 1989.
\newblock ISSN 1431-584X.
\newblock \doi{10.1007/BF01313666}.
\newblock URL \url{https://doi.org/10.1007/BF01313666}.

\bibitem[Mergenthaler and Pietralla(1994)]{Mergenthaler_ZPhysB}
D.~B. Mergenthaler and M.~Pietralla.
\newblock Heat conduction in highly oriented polyethylene.
\newblock \emph{Zeitschrift f{\"u}r Physik B Condensed Matter}, 94\penalty0
  (4):\penalty0 461--468, 1994.
\newblock \doi{10.1007/BF01317408}.
\newblock URL \url{https://doi.org/10.1007/BF01317408}.

\bibitem[Gibson et~al.(1977)Gibson, Greig, Sahota, Ward, and Choy]{Gibson_1977}
A.~G. Gibson, D.~Greig, M.~Sahota, I.~M. Ward, and C.~L. Choy.
\newblock Thermal conductivity of ultrahigh-modulus polyethylene.
\newblock \emph{Journal of Polymer Science: Polymer Letters Edition},
  15\penalty0 (4):\penalty0 183--192, 1977.
\newblock \doi{10.1002/pol.1977.130150401}.
\newblock URL
  \url{https://onlinelibrary.wiley.com/doi/abs/10.1002/pol.1977.130150401}.

\bibitem[Wang et~al.(2013)Wang, Ho, Segalman, and Cahill]{Wang_Mamo_2013}
Xiaojia Wang, Victor Ho, Rachel~A. Segalman, and David~G. Cahill.
\newblock Thermal conductivity of high-modulus polymer fibers.
\newblock \emph{Macromolecules}, 46\penalty0 (12):\penalty0 4937--4943, 2013.
\newblock \doi{10.1021/ma400612y}.
\newblock URL \url{https://doi.org/10.1021/ma400612y}.

\bibitem[Shen et~al.(2010)Shen, Henry, Tong, Zheng, and
  Chen]{Shen_NatNano_2010}
Sheng Shen, Asegun Henry, Jonathan Tong, Ruiting Zheng, and Gang Chen.
\newblock Polyethylene nanofibres with very high thermal conductivities.
\newblock \emph{Nature Nanotechnology}, 5\penalty0 (4):\penalty0 251--255,
  April 2010.
\newblock ISSN 1748-3395.
\newblock \doi{10.1038/nnano.2010.27}.
\newblock URL \url{https://www.nature.com/articles/nnano.2010.27}.

\bibitem[Shrestha et~al.(2018)Shrestha, Li, Chatterjee, Zheng, Wu, Liu, Luo,
  Choi, Hippalgaonkar, Boer, and Shen]{Shrestha_NatCommun_2018}
Ramesh Shrestha, Pengfei Li, Bikramjit Chatterjee, Teng Zheng, Xufei Wu, Zeyu
  Liu, Tengfei Luo, Sukwon Choi, Kedar Hippalgaonkar, Maarten P.~de Boer, and
  Sheng Shen.
\newblock Crystalline polymer nanofibers with ultra-high strength and thermal
  conductivity.
\newblock \emph{Nature Communications}, 9\penalty0 (1):\penalty0 1--9, April
  2018.
\newblock ISSN 2041-1723.
\newblock \doi{10.1038/s41467-018-03978-3}.
\newblock URL \url{https://www.nature.com/articles/s41467-018-03978-3}.

\bibitem[Rastogi et~al.(2011)Rastogi, Yao, Ronca, Bos, and van~der
  Eem]{Rastogi_Mamo_2011}
Sanjay Rastogi, Yefeng Yao, Sara Ronca, Johan Bos, and Joris van~der Eem.
\newblock Unprecedented high-modulus high-strength tapes and films of ultrahigh
  molecular weight polyethylene via solvent-free route.
\newblock \emph{Macromolecules}, 44\penalty0 (14):\penalty0 5558--5568, 2011.
\newblock \doi{10.1021/ma200667m}.
\newblock URL \url{https://doi.org/10.1021/ma200667m}.

\bibitem[Rastogi et~al.(2005)Rastogi, Lippits, Peters, Graf, Yao, and
  Spiess]{Rastogi_nmat}
Sanjay Rastogi, Dirk~R. Lippits, Gerrit W.~M. Peters, Robert Graf, Yefeng Yao,
  and Hans~W. Spiess.
\newblock Heterogeneity in polymer melts from melting of polymer crystals.
\newblock \emph{Nature Materials}, 4\penalty0 (8):\penalty0 635--641, Aug 2005.
\newblock ISSN 1476-4660.
\newblock \doi{10.1038/nmat1437}.
\newblock URL \url{https://doi.org/10.1038/nmat1437}.

\bibitem[Ronca et~al.(2017)Ronca, Igarashi, Forte, and
  Rastogi]{Ronca_Polymer_2017}
Sara Ronca, Tamito Igarashi, Giuseppe Forte, and Sanjay Rastogi.
\newblock Metallic-like thermal conductivity in a lightweight insulator:
  {Solid}-state processed {Ultra} {High} {Molecular} {Weight} {Polyethylene}
  tapes and films.
\newblock \emph{Polymer}, 123:\penalty0 203--210, August 2017.
\newblock ISSN 0032-3861.
\newblock \doi{10.1016/j.polymer.2017.07.027}.
\newblock URL
  \url{http://www.sciencedirect.com/science/article/pii/S0032386117306869}.

\bibitem[Xu et~al.(2019)Xu, Kraemer, Song, Jiang, Zhou, Loomis, Wang, Li,
  Ghasemi, Huang, Li, and Chen]{Xu_NatCommun_2019}
Yanfei Xu, Daniel Kraemer, Bai Song, Zhang Jiang, Jiawei Zhou, James Loomis,
  Jianjian Wang, Mingda Li, Hadi Ghasemi, Xiaopeng Huang, Xiaobo Li, and Gang
  Chen.
\newblock Nanostructured polymer films with metal-like thermal conductivity.
\newblock \emph{Nature Communications}, 10\penalty0 (1):\penalty0 1--8, April
  2019.
\newblock ISSN 2041-1723.
\newblock \doi{10.1038/s41467-019-09697-7}.
\newblock URL \url{https://www.nature.com/articles/s41467-019-09697-7}.

\bibitem[Lu et~al.(2017)Lu, Chiang, Du, Li, Gan, Zhang, Chu, Yao, Li, and
  Kang]{Lu_Polymer_2017}
Chenhao Lu, Sum~Wai Chiang, Hongda Du, Jia Li, Lin Gan, Xing Zhang, Xiaodong
  Chu, Youwei Yao, Baohua Li, and Feiyu Kang.
\newblock Thermal conductivity of electrospinning chain-aligned polyethylene
  oxide (peo).
\newblock \emph{Polymer}, 115:\penalty0 52 -- 59, 2017.
\newblock ISSN 0032-3861.
\newblock \doi{https://doi.org/10.1016/j.polymer.2017.02.024}.
\newblock URL
  \url{http://www.sciencedirect.com/science/article/pii/S0032386117301477}.

\bibitem[Singh et~al.(2014)Singh, Bougher, Weathers, Cai, Bi, Pettes,
  McMenamin, Lv, Resler, Gattuso, Altman, Sandhage, Shi, Henry, and
  Cola]{Singh_Natnano_2014}
Virendra Singh, Thomas~L. Bougher, Annie Weathers, Ye~Cai, Kedong Bi,
  Michael~T. Pettes, Sally~A. McMenamin, Wei Lv, Daniel~P. Resler, Todd~R.
  Gattuso, David~H. Altman, Kenneth~H. Sandhage, Li~Shi, Asegun Henry, and
  Baratunde~A. Cola.
\newblock High thermal conductivity of chain-oriented amorphous polythiophene.
\newblock \emph{Nature Nanotechnology}, 9\penalty0 (5):\penalty0 384--390,
  2014.
\newblock \doi{10.1038/nnano.2014.44}.
\newblock URL \url{https://doi.org/10.1038/nnano.2014.44}.

\bibitem[Bunn and Alcock(1945)]{Bunn_1945}
C.~W. Bunn and T.~C. Alcock.
\newblock The texture of polythene.
\newblock \emph{Trans. Faraday Soc.}, 41:\penalty0 317--325, 1945.
\newblock \doi{10.1039/TF9454100317}.
\newblock URL \url{http://dx.doi.org/10.1039/TF9454100317}.

\bibitem[Keller(1968)]{Keller_1968}
A~Keller.
\newblock Polymer crystals.
\newblock \emph{Reports on Progress in Physics}, 31\penalty0 (2):\penalty0
  623--704, jul 1968.
\newblock \doi{10.1088/0034-4885/31/2/304}.
\newblock URL \url{https://doi.org/10.1088/0034-4885/31/2/304}.

\bibitem[Wilson~III and Pake(1953)]{Wilson_NMR_1953}
C.~W. Wilson~III and G.~E. Pake.
\newblock Nuclear magnetic resonance determination of degree of crystallinity
  in two polymers.
\newblock \emph{Journal of Polymer Science}, 10\penalty0 (5):\penalty0
  503--505, 1953.
\newblock \doi{https://doi.org/10.1002/pol.1953.120100508}.
\newblock URL
  \url{https://onlinelibrary.wiley.com/doi/abs/10.1002/pol.1953.120100508}.

\bibitem[Peterlin and Meinel(1965)]{PeterlinMeinel_Calorimetry_1965}
A.~Peterlin and G.~Meinel.
\newblock Heat content of amorphous regions of drawn linear polyethylne.
\newblock \emph{Journal of Polymer Science Part B: Polymer Letters}, 3\penalty0
  (9):\penalty0 783--787, 1965.
\newblock \doi{https://doi.org/10.1002/pol.1965.110030919}.
\newblock URL
  \url{https://onlinelibrary.wiley.com/doi/abs/10.1002/pol.1965.110030919}.

\bibitem[Fischer and Schmidt(1962)]{Fischer_1962}
E.~W. Fischer and G.~F. Schmidt.
\newblock Long periods in drawn polyethylene.
\newblock \emph{Angewandte Chemie International Edition in English}, 1\penalty0
  (9):\penalty0 488--499, 1962.
\newblock \doi{https://doi.org/10.1002/anie.196204881}.
\newblock URL
  \url{https://onlinelibrary.wiley.com/doi/abs/10.1002/anie.196204881}.

\bibitem[KOBAYASHI and KUROKAWA(1962)]{KOBAYASHI_SAXS_Nat1962}
K.~KOBAYASHI and M.~KUROKAWA.
\newblock Small-angle diffraction of polyethylene.
\newblock \emph{Nature}, 196\penalty0 (4854):\penalty0 538--539, Nov 1962.
\newblock ISSN 1476-4687.
\newblock \doi{10.1038/196538a0}.
\newblock URL \url{https://doi.org/10.1038/196538a0}.

\bibitem[Yeh and Geil(1967)]{Yeh_1967}
G.~S.~Y. Yeh and P.~H. Geil.
\newblock Selected-area small-angle electron diffraction.
\newblock \emph{Journal of Materials Science}, 2\penalty0 (5):\penalty0
  457--469, Sep 1967.
\newblock ISSN 1573-4803.
\newblock \doi{10.1007/BF00562952}.
\newblock URL \url{https://doi.org/10.1007/BF00562952}.

\bibitem[Geil et~al.(1964)Geil, Anderson, Wunderlich, and Arakawa]{Geil_1964}
Phillip~H. Geil, Franklin~R. Anderson, Bernhard Wunderlich, and Tamio Arakawa.
\newblock Morphology of polyethylene crystallized from the melt under pressure.
\newblock \emph{Journal of Polymer Science Part A: General Papers}, 2\penalty0
  (8):\penalty0 3707--3720, 1964.
\newblock \doi{https://doi.org/10.1002/pol.1964.100020829}.
\newblock URL
  \url{https://onlinelibrary.wiley.com/doi/abs/10.1002/pol.1964.100020829}.

\bibitem[Peterlin(1971)]{Peterlin_theory_1971}
A.~Peterlin.
\newblock Molecular model of drawing polyethylene and polypropylene.
\newblock \emph{Journal of Materials Science}, 6\penalty0 (6):\penalty0
  490--508, Jun 1971.
\newblock ISSN 1573-4803.
\newblock \doi{10.1007/BF00550305}.
\newblock URL \url{https://doi.org/10.1007/BF00550305}.

\bibitem[Danner et~al.(1964)Danner, Boutin, and Safford]{Danner_INS_1964}
H.~R. Danner, H.~Boutin, and G.~J. Safford.
\newblock Low‐frequency molecular vibrations in solid hexane by neutron
  inelastic scattering.
\newblock \emph{The Journal of Chemical Physics}, 41\penalty0 (11):\penalty0
  3649--3650, 1964.
\newblock \doi{10.1063/1.1725784}.
\newblock URL \url{https://doi.org/10.1063/1.1725784}.

\bibitem[Safford and Naumann(1967)]{Safford_INS_1967}
G.~J. Safford and A.~W. Naumann.
\newblock Low frequency motions in polymers as measured by neutron inelastic
  scattering.
\newblock In \emph{Fortschritte der Hochpolymeren-Forschung}, pages 1--27,
  Berlin, Heidelberg, 1967. Springer Berlin Heidelberg.
\newblock ISBN 978-3-540-34908-2.

\bibitem[Holliday and White(1971)]{Holliday_INS_1971}
L.~Holliday and J.~W. White.
\newblock The stiffness of polymers in relation to their structure:.
\newblock \emph{Pure and Applied Chemistry}, 26\penalty0 (3-4):\penalty0
  545--582, 1971.
\newblock \doi{doi:10.1351/pac197126030545}.
\newblock URL \url{https://doi.org/10.1351/pac197126030545}.

\bibitem[Feldkamp et~al.(1968)Feldkamp, Venkataraman, and King]{Feldkamp1968}
L.~A. Feldkamp, G.~Venkataraman, and J.~S. King.
\newblock \emph{Dispersion Relation for Skeletal Vibrations in Deuterated
  Polyethylene}.
\newblock IAEA, International Atomic Energy Agency (IAEA), 1968.
\newblock URL \url{http://inis.iaea.org/search/search.aspx?orig_q=RN:44068946}.

\bibitem[Twisleton et~al.(1982)Twisleton, White, and
  Reynolds]{Twistleton_C11_11}
J.F Twisleton, J.W White, and P.A Reynolds.
\newblock Dynamical studies of fully oriented deuteropolyethylene by inelastic
  neturon scattering.
\newblock \emph{Polymer}, 23\penalty0 (4):\penalty0 578--588, 1982.
\newblock ISSN 0032-3861.
\newblock \doi{https://doi.org/10.1016/0032-3861(82)90097-0}.
\newblock URL
  \url{https://www.sciencedirect.com/science/article/pii/0032386182900970}.

\bibitem[Heyer et~al.(1984)Heyer, Buchenau, and Stamm]{Heyer_INS}
D.~Heyer, U.~Buchenau, and M.~Stamm.
\newblock Determination of elastic shear constants of polyethylene at room
  temperature by inelastic neutron scattering.
\newblock \emph{Journal of Polymer Science: Polymer Physics Edition},
  22\penalty0 (8):\penalty0 1515--1527, 1984.
\newblock \doi{https://doi.org/10.1002/pol.1984.180220814}.
\newblock URL
  \url{https://onlinelibrary.wiley.com/doi/abs/10.1002/pol.1984.180220814}.

\bibitem[Mermet et~al.(2003)Mermet, David, Lorenzen, and
  Krisch]{Mermet_IXS_2003}
A.~Mermet, L.~David, M.~Lorenzen, and M.~Krisch.
\newblock Inelastic x-ray scattering from stretch-oriented polyethylene.
\newblock \emph{The Journal of Chemical Physics}, 119\penalty0 (3):\penalty0
  1879--1884, 2003.
\newblock \doi{10.1063/1.1579681}.
\newblock URL \url{https://doi.org/10.1063/1.1579681}.

\bibitem[Smith et~al.(1985)Smith, Boudet, and Chanzy]{Smith_EM_1985}
Paul Smith, Alain Boudet, and Henri Chanzy.
\newblock The structure of ultradrawn high molecular weight polyethylene
  revealed by electron microscopy at 100 and 1500 {kV}.
\newblock \emph{Journal of Materials Science Letters}, 4\penalty0 (1):\penalty0
  13--18, January 1985.
\newblock ISSN 0261-8028, 1573-4811.
\newblock \doi{10.1007/BF00719883}.
\newblock URL \url{http://link.springer.com/10.1007/BF00719883}.

\bibitem[Brady and Thomas(1989)]{BRADY_TEM_1989}
Jean~M. Brady and Edwin~L. Thomas.
\newblock Conversion of single crystal mats to ultrahigh modulus polyethylene:
  the formation of a continuous crystalline phase.
\newblock \emph{Polymer}, 30\penalty0 (9):\penalty0 1615 -- 1622, 1989.
\newblock ISSN 0032-3861.
\newblock \doi{https://doi.org/10.1016/0032-3861(89)90320-0}.
\newblock URL
  \url{http://www.sciencedirect.com/science/article/pii/0032386189903200}.

\bibitem[Van~Aerle and Braam(1988)]{VanAerle_Xray_1988}
N.~A. J.~M. Van~Aerle and A.~W.~M. Braam.
\newblock A structural study on solid state drawing of solution-crystallized
  ultra-high molecular weight polyethylene.
\newblock \emph{Journal of Materials Science}, 23\penalty0 (12):\penalty0
  4429--4436, 1988.
\newblock \doi{10.1007/BF00551941}.
\newblock URL \url{https://doi.org/10.1007/BF00551941}.

\bibitem[Hu and Schmidt-Rohr(2000)]{Hu_NMR_2000}
W.-G Hu and K~Schmidt-Rohr.
\newblock Characterization of ultradrawn polyethylene fibers by nmr:
  crystallinity, domain sizes and a highly mobile second amorphous phase.
\newblock \emph{Polymer}, 41\penalty0 (8):\penalty0 2979 -- 2987, 2000.
\newblock ISSN 0032-3861.
\newblock \doi{https://doi.org/10.1016/S0032-3861(99)00429-2}.
\newblock URL
  \url{http://www.sciencedirect.com/science/article/pii/S0032386199004292}.

\bibitem[Litvinov et~al.(2011)Litvinov, Xu, Melian, Demco, Möller, and
  Simmelink]{Litvinov_MaMo_2011}
V.~M. Litvinov, Jianjun Xu, C.~Melian, D.~E. Demco, M.~Möller, and
  J.~Simmelink.
\newblock Morphology, chain dynamics, and domain sizes in highly drawn gel-spun
  ultrahigh molecular weight polyethylene fibers at the final stages of drawing
  by saxs, waxs, and 1h solid-state nmr.
\newblock \emph{Macromolecules}, 44\penalty0 (23):\penalty0 9254--9266, 2011.
\newblock \doi{10.1021/ma201888f}.
\newblock URL \url{https://doi.org/10.1021/ma201888f}.

\bibitem[Magonov et~al.(1993)Magonov, Sheiko, Deblieck, and
  Moller]{Magonov_AFM_1993}
S.~N. Magonov, S.~S. Sheiko, R.~A.~C. Deblieck, and M.~Moller.
\newblock Atomic-force microscopy of gel-drawn ultrahigh-molecular-weight
  polyethylene.
\newblock \emph{Macromolecules}, 26\penalty0 (6):\penalty0 1380--1386, 1993.
\newblock \doi{10.1021/ma00058a029}.
\newblock URL \url{https://doi.org/10.1021/ma00058a029}.

\bibitem[Zubov et~al.(1992)Zubov, Chvalun, Selikhova, Konstantinopolskaya, and
  Bakeev]{Zubov_SAXS_1992}
Y.~A. Zubov, S.~N. Chvalun, V.~I. Selikhova, M.~B. Konstantinopolskaya, and
  N.~Ph. Bakeev.
\newblock The structure of highly oriented high modulus polyethylene.
\newblock \emph{Polymer Engineering \& Science}, 32\penalty0 (17):\penalty0
  1316--1324, 1992.
\newblock \doi{10.1002/pen.760321720}.
\newblock URL
  \url{https://onlinelibrary.wiley.com/doi/abs/10.1002/pen.760321720}.

\bibitem[Anandakumaran et~al.(1988)Anandakumaran, Roy, and
  Manley]{Anandakumaran_Mamo_1988}
K.~Anandakumaran, S.~K. Roy, and R.~St.~John Manley.
\newblock Drawing-induced changes in the properties of polyethylene fibers
  prepared by gelation/crystallization.
\newblock \emph{Macromolecules}, 21\penalty0 (6):\penalty0 1746--1751, 1988.
\newblock \doi{10.1021/ma00184a036}.
\newblock URL \url{https://doi.org/10.1021/ma00184a036}.

\bibitem[Stein and Norris(1956)]{Stein_xray}
Richard~S. Stein and Forrest~H. Norris.
\newblock The x-ray diffraction, birefringence, and infrared dichroism of
  stretched polyethylene.
\newblock \emph{Journal of Polymer Science}, 21\penalty0 (99):\penalty0
  381--396, 1956.
\newblock \doi{https://doi.org/10.1002/pol.1956.120219903}.
\newblock URL
  \url{https://onlinelibrary.wiley.com/doi/abs/10.1002/pol.1956.120219903}.

\bibitem[Tang et~al.(2007)Tang, Jiang, Men, An, Enderle, Lilge, Roth, Gehrke,
  and Rieger]{TANG_SAXS}
Yujing Tang, Zhiyong Jiang, Yongfeng Men, Lijia An, Hans-Friedrich Enderle,
  Dieter Lilge, Stephan~V. Roth, Rainer Gehrke, and Jens Rieger.
\newblock Uniaxial deformation of overstretched polyethylene: In-situ
  synchrotron small angle x-ray scattering study.
\newblock \emph{Polymer}, 48\penalty0 (17):\penalty0 5125 -- 5132, 2007.
\newblock ISSN 0032-3861.
\newblock \doi{https://doi.org/10.1016/j.polymer.2007.06.056}.
\newblock URL
  \url{http://www.sciencedirect.com/science/article/pii/S0032386107006313}.

\bibitem[Choy et~al.(1993)Choy, Fei, and Xi]{Choy_Gelspun_1993}
C.~L. Choy, Y.~Fei, and T.~G. Xi.
\newblock Thermal conductivity of gel-spun polyethylene fibers.
\newblock \emph{Journal of Polymer Science Part B: Polymer Physics},
  31\penalty0 (3):\penalty0 365--370, 1993.
\newblock \doi{10.1002/polb.1993.090310315}.
\newblock URL
  \url{https://onlinelibrary.wiley.com/doi/abs/10.1002/polb.1993.090310315}.

\bibitem[Hennig(1967)]{Hennig}
Jürgen Hennig.
\newblock Anisotropy and structure in uniaxially stretched amorphous high
  polymers.
\newblock \emph{Journal of Polymer Science Part C: Polymer Symposia},
  16\penalty0 (5):\penalty0 2751--2761, 1967.
\newblock \doi{10.1002/polc.5070160528}.
\newblock URL
  \url{https://onlinelibrary.wiley.com/doi/abs/10.1002/polc.5070160528}.

\bibitem[Choy and Young(1977)]{ChoyandYoung_1977}
C.L. Choy and K.~Young.
\newblock Thermal conductivity of semicrystalline polymers — a model.
\newblock \emph{Polymer}, 18\penalty0 (8):\penalty0 769--776, 1977.
\newblock ISSN 0032-3861.
\newblock \doi{https://doi.org/10.1016/0032-3861(77)90179-3}.
\newblock URL
  \url{https://www.sciencedirect.com/science/article/pii/0032386177901793}.

\bibitem[Takayanagi et~al.(1964)Takayanagi, Uemura, and Minami]{Takayanagi}
Motowo Takayanagi, Shinsaku Uemura, and Shunsuke Minami.
\newblock Application of equivalent model method to dynamic rheo-optical
  properties of crystalline polymer.
\newblock \emph{Journal of Polymer Science Part C: Polymer Symposia},
  5\penalty0 (1):\penalty0 113--122, 1964.
\newblock \doi{10.1002/polc.5070050111}.
\newblock URL
  \url{https://onlinelibrary.wiley.com/doi/abs/10.1002/polc.5070050111}.

\bibitem[Robbins et~al.(2019)Robbins, Drakopoulos, Martin-Fabiani, Ronca, and
  Minnich]{ABR_PNAS_2019}
Andrew~B. Robbins, Stavros~X. Drakopoulos, Ignacio Martin-Fabiani, Sara Ronca,
  and Austin~J. Minnich.
\newblock Ballistic thermal phonons traversing nanocrystalline domains in
  oriented polyethylene.
\newblock \emph{Proceedings of the National Academy of Sciences}, 116\penalty0
  (35):\penalty0 17163--17168, August 2019.
\newblock ISSN 0027-8424, 1091-6490.
\newblock \doi{10.1073/pnas.1905492116}.
\newblock URL \url{https://www.pnas.org/content/116/35/17163}.

\bibitem[Drakopoulos et~al.(2020)Drakopoulos, Tarallo, Guan, Martin-Fabiani,
  and Ronca]{Stavros_AuSize}
Stavros~X. Drakopoulos, Oreste Tarallo, Linlin Guan, Ignacio Martin-Fabiani,
  and Sara Ronca.
\newblock Nanocomposites of au/disentangled uhmwpe: A combined optical and
  structural study.
\newblock \emph{Molecules}, 25\penalty0 (14), 2020.
\newblock ISSN 1420-3049.
\newblock \doi{10.3390/molecules25143225}.
\newblock URL \url{https://www.mdpi.com/1420-3049/25/14/3225}.

\bibitem[Johnson et~al.(2013)Johnson, Maznev, Cuffe, Eliason, Minnich, Kehoe,
  Torres, Chen, and Nelson]{Jeremy_PRL_2013}
Jeremy~A. Johnson, A.~A. Maznev, John Cuffe, Jeffrey~K. Eliason, Austin~J.
  Minnich, Timothy Kehoe, Clivia M.~Sotomayor Torres, Gang Chen, and Keith~A.
  Nelson.
\newblock Direct measurement of room-temperature nondiffusive thermal transport
  over micron distances in a silicon membrane.
\newblock \emph{Phys. Rev. Lett.}, 110:\penalty0 025901, Jan 2013.
\newblock \doi{10.1103/PhysRevLett.110.025901}.
\newblock URL \url{https://link.aps.org/doi/10.1103/PhysRevLett.110.025901}.

\bibitem[Chang(1974)]{Chang_heatcap}
SS~Chang.
\newblock Heat capacities of polyethylene from 2 to 360 k. ii. two high density
  linear polyethylene samples and thermodynamic properties of crystalline
  linear polyethylene.
\newblock \emph{J. Res. NBS A Phys. Chem}, 3, 1974.

\bibitem[Choy(1977)]{Choy_polymer_1977}
C.L. Choy.
\newblock Thermal conductivity of polymers.
\newblock \emph{Polymer}, 18\penalty0 (10):\penalty0 984 -- 1004, 1977.
\newblock ISSN 0032-3861.
\newblock \doi{https://doi.org/10.1016/0032-3861(77)90002-7}.
\newblock URL
  \url{http://www.sciencedirect.com/science/article/pii/0032386177900027}.

\bibitem[Zolotoyabko(2011)]{Zolotoyabko_Book}
E.~Zolotoyabko.
\newblock \emph{Basic Concepts of Crystallography}.
\newblock Wiley, 2011.
\newblock ISBN 9783527330096.
\newblock URL \url{https://books.google.com/books?id=eNdJYgEACAAJ}.
\newblock in page 195.

\bibitem[Nysten et~al.(1995)Nysten, Gonry, and Issi]{NYSTEN_1995}
B.~Nysten, P.~Gonry, and J.-P. Issi.
\newblock Intra-and interchain thermal conduction in polymers.
\newblock \emph{Synthetic Metals}, 69\penalty0 (1):\penalty0 67--68, 1995.
\newblock ISSN 0379-6779.
\newblock \doi{https://doi.org/10.1016/0379-6779(94)02366-7}.
\newblock URL
  \url{https://www.sciencedirect.com/science/article/pii/0379677994023667}.
\newblock Proceedings of the International Conference on Science and Technology
  of Synthetic Metals.

\bibitem[Fujishiro et~al.(1998)Fujishiro, Ikebe, Kashima, and
  Yamanaka]{Fujishiro_1998}
Hiroyuki Fujishiro, Manabu Ikebe, Toshihiro Kashima, and Atsuhiko Yamanaka.
\newblock Drawing effect on thermal properties of high-strength polyethylene
  fibers.
\newblock \emph{Japanese Journal of Applied Physics}, 37\penalty0 (Part 1, No.
  4A):\penalty0 1994--1995, apr 1998.
\newblock \doi{10.1143/jjap.37.1994}.
\newblock URL \url{https://doi.org/10.1143/jjap.37.1994}.

\bibitem[Minnich(2012)]{Austin_Determining}
A.~J. Minnich.
\newblock Determining phonon mean free paths from observations of
  quasiballistic thermal transport.
\newblock \emph{Phys. Rev. Lett.}, 109:\penalty0 205901, Nov 2012.
\newblock \doi{10.1103/PhysRevLett.109.205901}.
\newblock URL \url{https://link.aps.org/doi/10.1103/PhysRevLett.109.205901}.

\bibitem[Ravichandran et~al.(2018)Ravichandran, Zhang, and
  Minnich]{Navaneeth_PRX_2018}
Navaneetha~K. Ravichandran, Hang Zhang, and Austin~J. Minnich.
\newblock Spectrally resolved specular reflections of thermal phonons from
  atomically rough surfaces.
\newblock \emph{Phys. Rev. X}, 8:\penalty0 041004, Oct 2018.
\newblock \doi{10.1103/PhysRevX.8.041004}.
\newblock URL \url{https://link.aps.org/doi/10.1103/PhysRevX.8.041004}.

\bibitem[Pietralla(1996)]{Pietralla1996}
M.~Pietralla.
\newblock High thermal conductivity of polymers: Possibility or dream?
\newblock \emph{Journal of Computer-Aided Materials Design}, 3\penalty0
  (1):\penalty0 273--280, Aug 1996.
\newblock ISSN 1573-4900.
\newblock \doi{10.1007/BF01185664}.
\newblock URL \url{https://doi.org/10.1007/BF01185664}.

\bibitem[Bowman and Krumhansl(1958)]{Bowman_elasticcont1958}
J.C. Bowman and J.A. Krumhansl.
\newblock The low-temperature specific heat of graphite.
\newblock \emph{Journal of Physics and Chemistry of Solids}, 6\penalty0
  (4):\penalty0 367--379, 1958.
\newblock ISSN 0022-3697.
\newblock \doi{https://doi.org/10.1016/0022-3697(58)90055-6}.
\newblock URL
  \url{https://www.sciencedirect.com/science/article/pii/0022369758900556}.

\bibitem[Chen et~al.(2013)Chen, Wei, Chen, and Dames]{Dames_PRB}
Z.~Chen, Z.~Wei, Y.~Chen, and C.~Dames.
\newblock Anisotropic debye model for the thermal boundary conductance.
\newblock \emph{Phys. Rev. B}, 87:\penalty0 125426, Mar 2013.
\newblock \doi{10.1103/PhysRevB.87.125426}.
\newblock URL \url{https://link.aps.org/doi/10.1103/PhysRevB.87.125426}.

\bibitem[Minnich(2015)]{Austin_layaniso}
A.~J. Minnich.
\newblock Phonon heat conduction in layered anisotropic crystals.
\newblock \emph{Phys. Rev. B}, 91:\penalty0 085206, Feb 2015.
\newblock \doi{10.1103/PhysRevB.91.085206}.
\newblock URL \url{https://link.aps.org/doi/10.1103/PhysRevB.91.085206}.

\bibitem[Shulumba et~al.(2017)Shulumba, Hellman, and Minnich]{Nina_PRL_2017}
Nina Shulumba, Olle Hellman, and Austin~J. Minnich.
\newblock Lattice thermal conductivity of polyethylene molecular crystals from
  first-principles including nuclear quantum effects.
\newblock \emph{Phys. Rev. Lett.}, 119:\penalty0 185901, Oct 2017.
\newblock \doi{10.1103/PhysRevLett.119.185901}.
\newblock URL \url{https://link.aps.org/doi/10.1103/PhysRevLett.119.185901}.

\bibitem[Zeller and Pohl(1971)]{Zeller_DPA}
R.~C. Zeller and R.~O. Pohl.
\newblock Thermal conductivity and specific heat of noncrystalline solids.
\newblock \emph{Phys. Rev. B}, 4:\penalty0 2029--2041, Sep 1971.
\newblock \doi{10.1103/PhysRevB.4.2029}.
\newblock URL \url{https://link.aps.org/doi/10.1103/PhysRevB.4.2029}.

\end{thebibliography}


\begin{thebibliography}{3}
\providecommand{\natexlab}[1]{#1}
\providecommand{\url}[1]{\texttt{#1}}
\expandafter\ifx\csname urlstyle\endcsname\relax
  \providecommand{\doi}[1]{doi: #1}\else
  \providecommand{\doi}{doi: \begingroup \urlstyle{rm}\Url}\fi

\bibitem[Fujishiro et~al.(1994)Fujishiro, Ikebe, Naito, Noto, Kohayashi, and
  Yoshizawa]{Fujishiro_1994}
Hiroyuki Fujishiro, Manabu Ikebe, Tomoyuki Naito, Koshichi Noto, Shuichi
  Kohayashi, and Shuji Yoshizawa.
\newblock Anisotropic thermal diffusivity and conductivity of {YBCO}(123) and
  {YBCO}(211) mixed crystals. i.
\newblock \emph{Japanese Journal of Applied Physics}, 33\penalty0 (Part 1, No.
  9A):\penalty0 4965--4970, sep 1994.
\newblock \doi{10.1143/jjap.33.4965}.
\newblock URL \url{https://doi.org/10.1143/jjap.33.4965}.

\bibitem[Burgess and Greig(1974)]{BurgessGreig_1974}
S~Burgess and D~Greig.
\newblock The low-temperature thermal conductivity of two-phase amorphous
  polymers.
\newblock \emph{Journal of Physics D: Applied Physics}, 7\penalty0
  (15):\penalty0 2051--2057, oct 1974.
\newblock \doi{10.1088/0022-3727/7/15/309}.
\newblock URL \url{https://doi.org/10.1088/0022-3727/7/15/309}.

\bibitem[Xu et~al.(2019)Xu, Kraemer, Song, Jiang, Zhou, Loomis, Wang, Li,
  Ghasemi, Huang, Li, and Chen]{Xu_NatCommun_2019}
Yanfei Xu, Daniel Kraemer, Bai Song, Zhang Jiang, Jiawei Zhou, James Loomis,
  Jianjian Wang, Mingda Li, Hadi Ghasemi, Xiaopeng Huang, Xiaobo Li, and Gang
  Chen.
\newblock Nanostructured polymer films with metal-like thermal conductivity.
\newblock \emph{Nature Communications}, 10\penalty0 (1):\penalty0 1--8, April
  2019.
\newblock ISSN 2041-1723.
\newblock \doi{10.1038/s41467-019-09697-7}.
\newblock URL \url{https://www.nature.com/articles/s41467-019-09697-7}.

\end{thebibliography}

\end{document}

\title{Supporting information: Origin of high thermal conductivity in disentangled ultra-high molecular weight polyethylene films: ballistic phonons within enlarged crystals}

\author{Taeyong Kim~\orcidicon{0000-0003-2452-1065}\,}

\affiliation{%
Division of Engineering and Applied Science, California Institute of Technology, Pasadena, California 91125, USA
}%

\author{Stavros X. Drakopoulos~\orcidicon{0000-0002-6798-0790}\,}
\affiliation{%
Department of Materials, Loughborough University, Loughborough LE11 3TU, United Kingdom
}%

\author{Sara Ronca~\orcidicon{0000-0003-3434-6352}\,}
\affiliation{%
Department of Materials, Loughborough University, Loughborough LE11 3TU, United Kingdom
}%

\author{Austin J. Minnich~\orcidicon{0000-0002-9671-9540} }
 \email{aminnich@caltech.edu}
\affiliation{%
Division of Engineering and Applied Science, California Institute of Technology, Pasadena, California 91125, USA
}%     

\date{\today}
{    \global\let\newpagegood\newpage
    \global\let\newpage\relax
\maketitle}

\clearpage

\section{Additional transient grating data}

Additional TG data for various grating periods and temperatures are presented below.

\begin{figure}[hbt!] 
\includegraphics[width=\textwidth,height=.4\textheight,keepaspectratio]{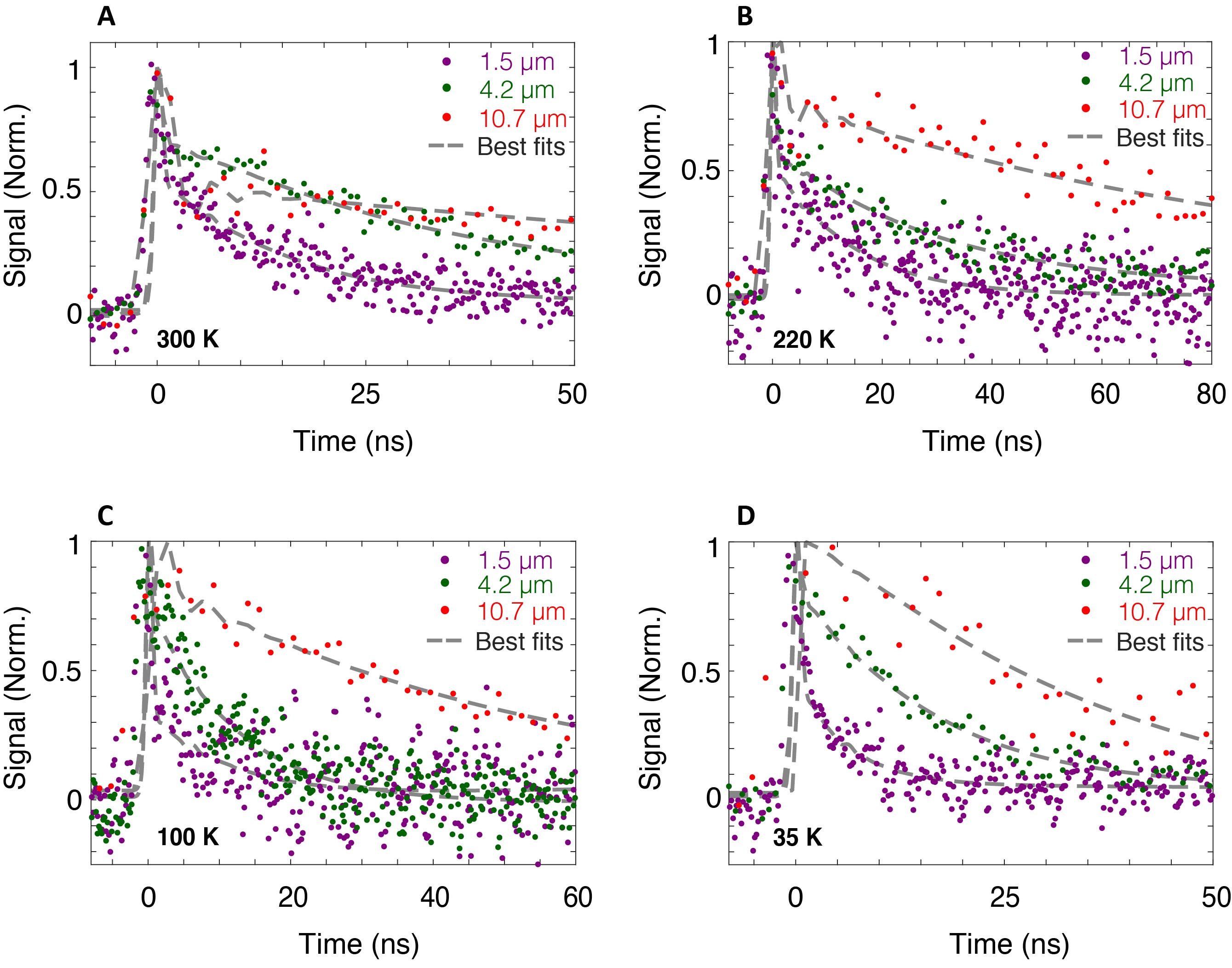}
\caption{Additional  TG measurements and  corresponding best fits for grating periods  of 1.5 $\mu m$, 4.2 $\mu m$, and 10.7 $\mu m$ at (A) 300 K (B) 220 K (C) 100 K (D) 35 K. }
\label{SIfig:TGRawsig}
\end{figure}
\clearpage

\section{Thermal conductivity versus  grating period}

\begin{figure}[hbt!] 
\includegraphics[width=\textwidth,height=\textheight/2,keepaspectratio]{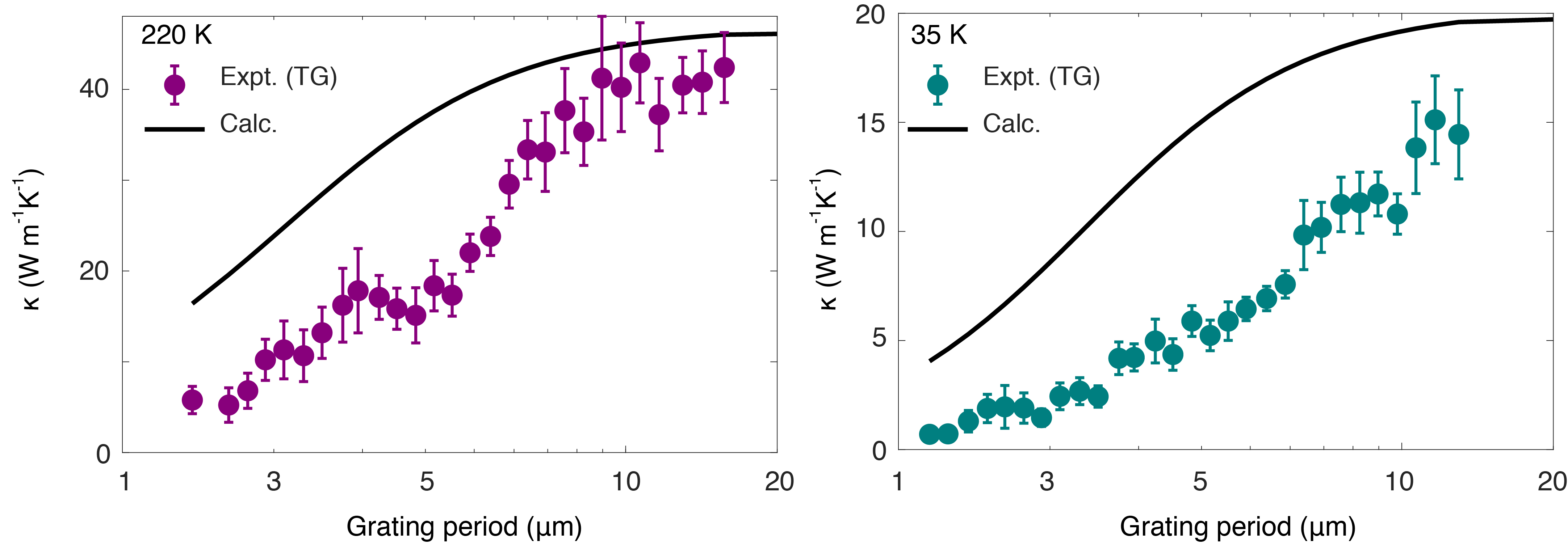}
\caption{Measured thermal conductivity versus  grating period along with the fit based on an anistropic Debye model at (A) 220 K and (B) 35 K.}
\label{SIfig:kvsGP}
\end{figure}

Additional measurements of thermal conductivity versus  grating period at 220 K and 35 K are shown in Fig.~\ref{SIfig:kvsGP} along with the model predictions. The model captures the general trend although quantitative discrepancies remain.

\clearpage

\section{Computed thermal conductivity versus grating period versus temperature using other candidate profiles}

\begin{figure}[hbt!] 
\includegraphics[width=\textwidth,height=\textheight/2,keepaspectratio]{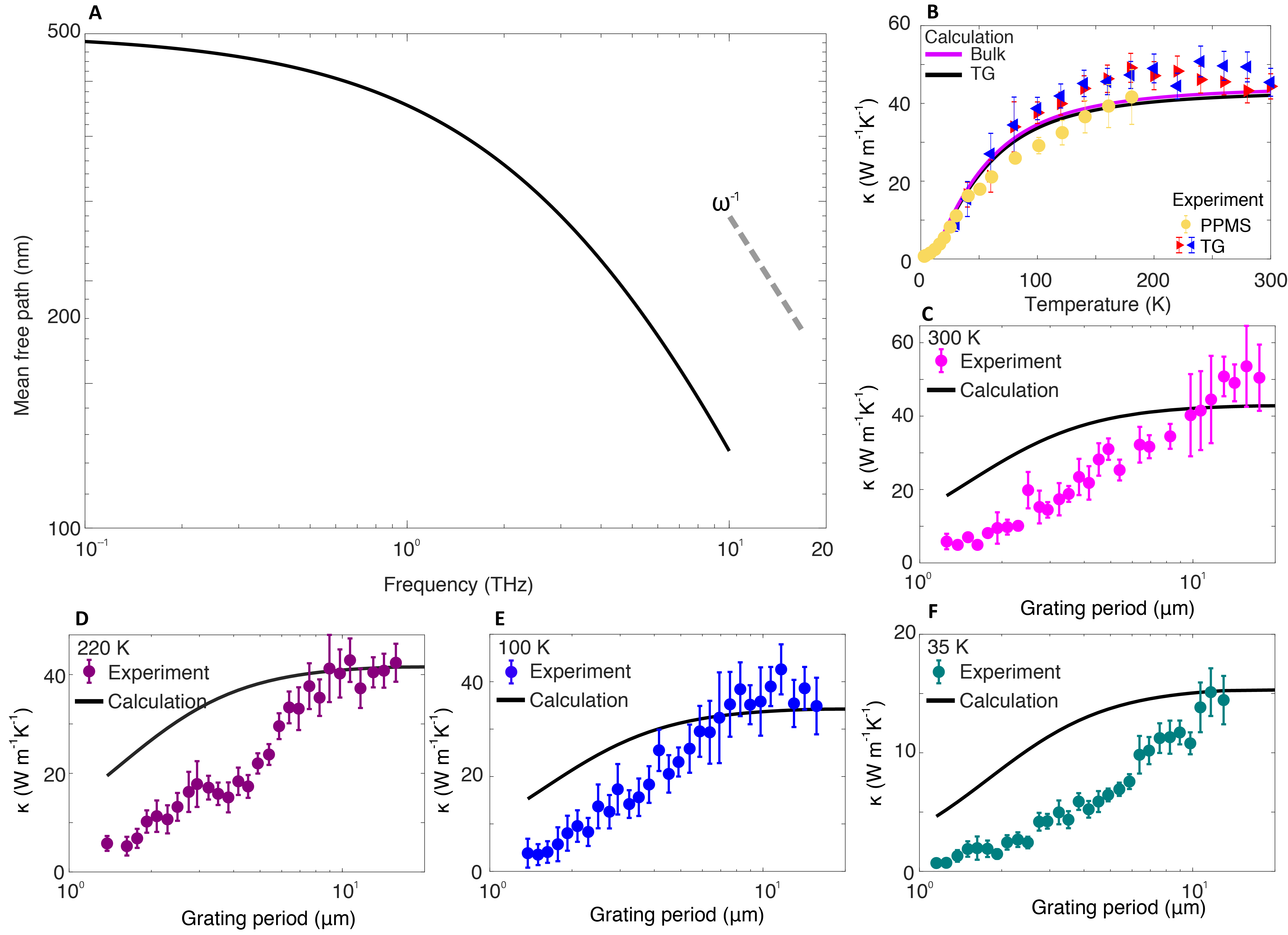}
\caption{(A) Candidate MFP profile versus frequency (Constant value (400nm) to $\omega^{-1}$, transition frequency $\sim 7$ THz). (B) Calculated bulk thermal conductivity versus temperature using the profile in (A). Calculated  thermal conductivity versus grating period at (C) 300 K, (D) 220 K, (E) 100 K, and (F) 35 K using the profile in (A).}
\label{SIfig:fitw1}
\end{figure}

\begin{figure}[hbt!] 
\includegraphics[width=\textwidth,height=\textheight/2,keepaspectratio]{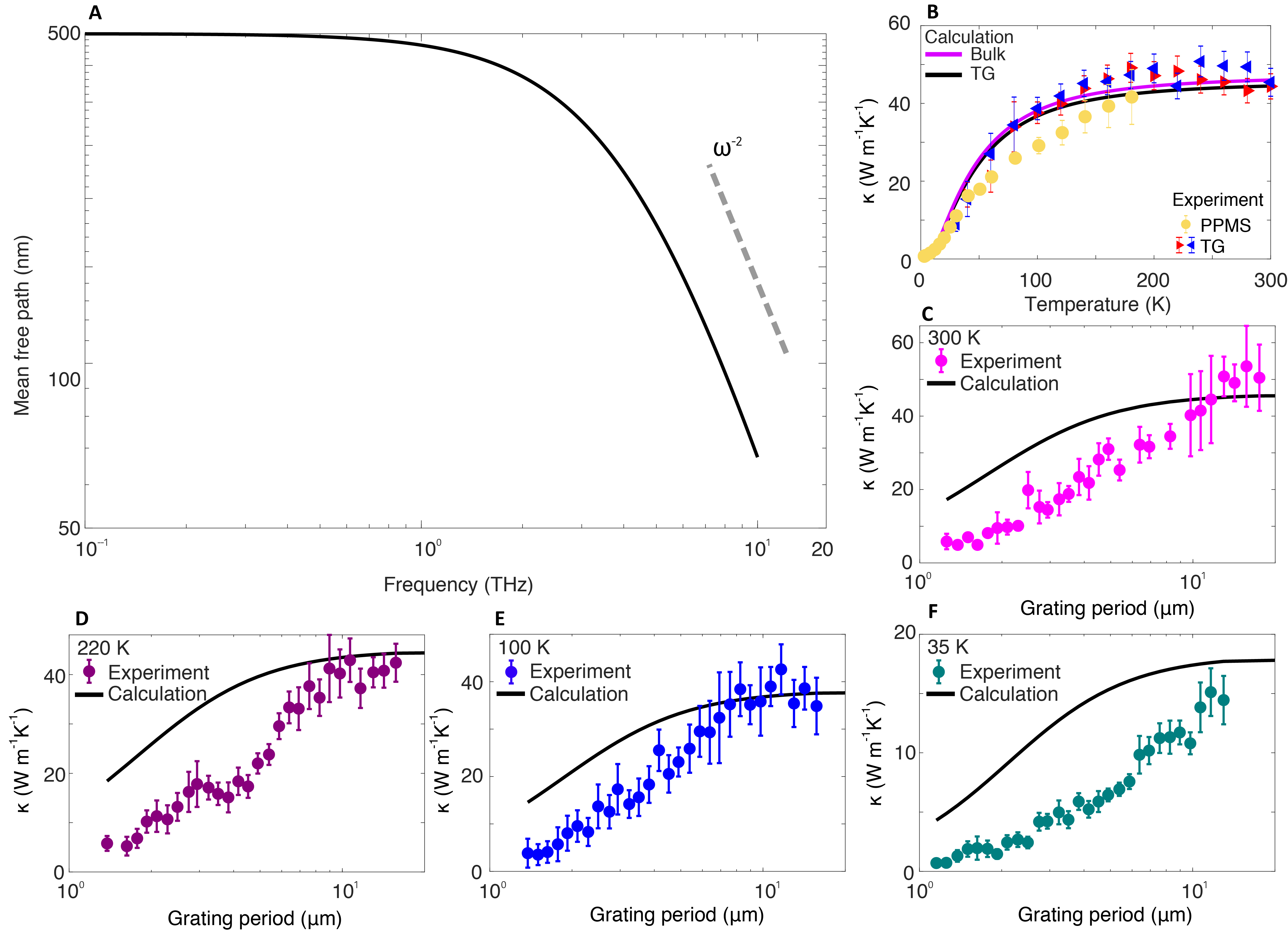}
\caption{(A) Candidate MFP profile versus frequency (Constant value (400nm) to $\omega^{-2}$, transition frequency $\sim 7$ THz).  (B) Calculated bulk thermal conductivity versus temperature using the profile in (A). Calculated thermal conductivity versus grating period at (C) 300 K, (D) 220 K, (E) 100 K, and (F) 35 K using the profile in (A).}
\label{SIfig:fitw2}
\end{figure}

This section presents calculated thermal conductivity using alternate  trends for the LA branch relaxation time versus frequency. The first trend is a constant MFP of 400 nm that transitions to  $\omega^{-1}$ at 7 THz, shown in Fig.~\ref{SIfig:fitw1}; the second is a constant MFP of 400 nm that transitions to $\omega^{-2}$ at 7 THz, shown in Fig.~\ref{SIfig:fitw2}. Power laws with larger exponents generally exhibit improved agreement of thermal conductivity versus grating period.

\clearpage

\section{Details of the Physical Property Measurement System measurements}

We performed bulk thermal conductivity measurements at cryogenic temperatures using a commercial 7 T Dynacool Physical Property Measurement System (PPMS, Quantum Design). Samples of 196 DR (thickness: 30 $\mu$m, as measured using the calipers and cross-sectional view from scanning electron microscopy; lateral width: 1.62 mm; heat conduction length excluding electrical contact: 5.6 mm) were mounted in a four-point electrical contact geometry. Silver conducting epoxy was applied between the sample and the four copper wires and cured for 7 hours at $\sim400$ K on a hot plate. Following Refs.~\cite{Fujishiro_1994,BurgessGreig_1974}, we additionally applied a cryogenic varnish on top of the epoxy (GE7031, Lakeshore) to ensure both the physical and the thermal contact between the wires and the sample; the varnish was cured at room temperature for 24 hours and then  transferred onto a thermal transport platform.

PPMS measurements were conducted at temperatures ranging from $\sim3$ - 200 K. At each temperature, we  adjusted the electrical heater power to minimize the temperature rise and temperature fluctuations during thermal equilibration. Table~\ref{tab:optpropaSi}, shows the base temperature and peak temperature  along with the electrical power. The temperature rise was constrained to be from 1.3 - 3.5\% of the base temperature. To account for radiative heat loss, we used an emissivity of 0.1 as given in Ref.~\cite{Xu_NatCommun_2019}.

\begin{table}[h] \centering
 \caption{\label{tab:optpropaSi}
Measured sample temperature and electrical power used in PPMS. }

 \begin{tabular}{cccc}
  \hline
   \hline
 Base sample temperature ($K$) & Heater power  ($\mu$W) &Peak temperature ($K$) \\ 
 \hline
 3  & 0.3 & 3.0\\
 3  & 0.5 & 3.1\\
 4  & 0.6 & 4.1\\
 8  & 4 & 8.3\\
 12 & 8 & 12.4\\
 16 & 20 & 16.7\\
 20 & 30 & 20.7\\
 25 & 50 & 25.8\\
 30 & 120   & 31.5\\
 41 & 240             & 42.0\\
 50 & 320             & 52.4\\
 61 & 400              & 62.5\\
 81 & 700              & 83.5\\
101 & 950             & 104.1\\
121 & $1.2\times10^{3}$              & 124.4 \\
141 & $1.05\times10^{3}$             & 143.2\\
161 & $1.45\times10^{3}$           & 163.9\\
181 & $1.48\times10^{3}$             &  183.5\\
   \hline
    \hline
\end{tabular}
\end{table}
\clearpage

\section{Steady heating at 30 K}

We estimate the steady heating due to the pump and probe pulses at 30 K. At higher temperatures, the thermal conductivity of the sample is high enough that the steady temperature rise relative to the base temperature can be neglected. Given the laser beam diameter $d = 550$ $\mu$m, and heat conduction length $l = 10$ mm (see Fig. 1B), the steady temperature rise considering 1D heat conduction along the drawing direction can be expressed as

\begin{eqnarray}
    \Delta T = \frac{l}{A} \frac{P_{abs,total}}{\kappa}
    \label{eq:steady30K}
\end{eqnarray}

where $A = d t \simeq 550$ $\mu$m $\times 30$ $\mu$m  = $1.7 \times 10^{-8}$ m$^2$ is the cross-sectional area, $P_{abs,total} = \alpha P_{total}$ is the absorbed average power  with incident optical power $P_{total}$, and $\kappa$ is the thermal conductivity at 30 K ($\sim10 $ $Wm^{-1}K^{-1}$). $P_{total}$ was calculated to be $\sim 1.75$ mW, using $P_{total} = (1/2)(E_{pump}\times f_{rep} + P_{probe})$ where $E_{pump}$ is incident pump energy ($\sim 13$ $\mu$J), $f_{rep}$ is laser repetition rate (200 Hz), and $P_{probe}$ is the steady incident probe power ($\approx 900$ $\mu$W). The factor of 1/2 accounts for heat conduction in both directions from the center of the film to the edges. Since experimentally determining $\alpha$ of the sample is challenging due to intense optical scattering of the sample, $\alpha$ was roughly estimated as $\sim 5$\%.  The resulting steady temperature rise is estimated to be $\Delta T \sim 10$~mm~$/(1.7 \times 10^{-8}$~m$^2$) $\times (1.75$ mW $\times 0.05 / 10 $ Wm$^{-1}$K$^{-1}$) $\sim 5.3$ K. Therefore, we take the temperature of the sample for a base temperature of 30 K to be $\sim 35$ K.

\clearpage

 \section{Surface characterization using atomic force microscopy (AFM)}

\begin{figure}[hbt!] 
{\includegraphics[width=\textwidth,height=\textheight/3,keepaspectratio]{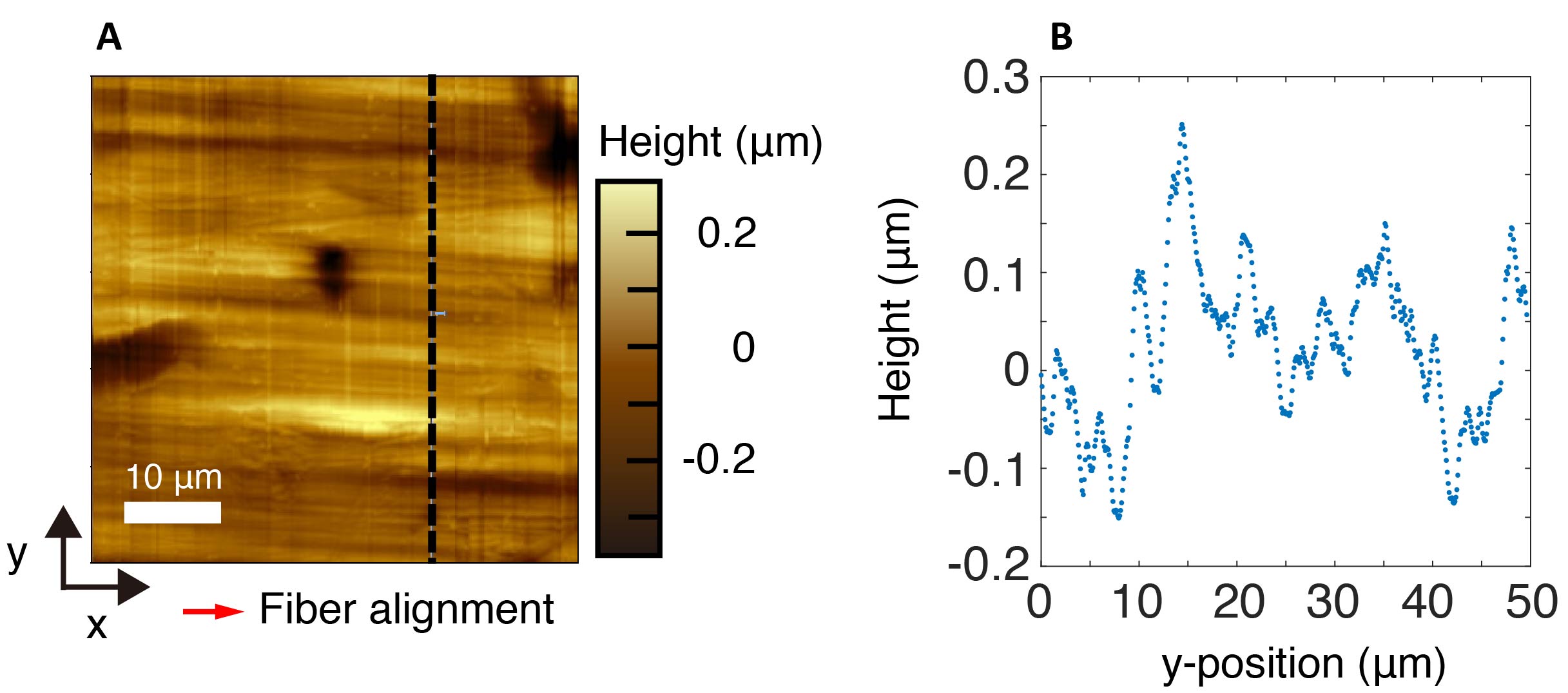}
\phantomsubcaption\label{fig:AFMtopo}
\phantomsubcaption\label{fig:AFMprofile}
}
\caption{AFM Characterization of the UHMWPE surface. A. AFM topographic image of the sample. The x- and y- direction indicates parallel (perpendicular) to the fiber alignment. B. AFM crosscut along the dashed line in (A). RMS roughness is calculated to be $\sim 70$ nm, and the maximum peak-to-valley difference is $\sim 360$  nm.}
\label{SIfig:AFM}
\end{figure}

This section provides the surface profile measured from atomic force microscopy (AFM). The AFM topography is shown in Fig.~\ref{SIfig:AFM}. The AFM crosscut perpendicular to the fiber alignment direction (dashed line in Fig.~\ref{fig:AFMtopo}) is shown in Fig.~\ref{fig:AFMprofile}. Calculated RMS roughness is $\sim 70$ nm, and the maximum peak-to-valley difference is $\sim 360$ nm. The height difference indicates the surface inhomogeneity over length scales comparable to the optical wavelength (515 nm) used, explaining the intense optical scattering observed in the TG experiment.

\clearpage

\bibliographystyle{unsrtnat}
\bibliography{PolymerRef.bib}